\documentstyle[equations,fleqn,epsf]{mn}

\newif\ifAMStwofonts

\newcommand{\beq}	{\begin{equation}}
\newcommand{\eeq}	{\end{equation}}
\newcommand{\bsq}	{\begin{subequations}}
\newcommand{\esq}	{\end{subequations}}
\newcommand{\bec}	{\begin{cases}}			
\newcommand{\eec}	{\end{cases}}			
\newcommand{\bea}	{\begin{array}}			
\newcommand{\eea}	{\end{array}}			
\newcommand{\ben}	{\begin{eqnarray}}
\newcommand{\een}	{\end{eqnarray}}
\newcommand{\benn}	{\begin{eqnarray*}}
\newcommand{\eenn}	{\end{eqnarray*}}

\renewcommand{\d}	{{\rm d}}			

\newcommand{\ds}	{\displaystyle}
\newcommand{\ts}	{\textstyle}
\newcommand{\fracj}[2]	{{\ts{#1\over#2}}}
\newcommand{\half}	{\fracj12}			
\newcommand{\e}		{{\rm e}}

\newcommand{\nc}[1]     {{\,{#1}\,}}			
\newcommand{\eq}        {\nc{=}}			
\newcommand{\id}	{\nc{\equiv}}			
\newcommand{\ap}	{\nc{\approx}}			
\newcommand{\mi}	{\nc{-}}			
\newcommand{\pl}	{\nc{+}}			

\newcommand{\EJ}	{{E_{\rm J}}}
\newcommand{\Lc}	{{L_{\rm c}}}
\newcommand{\Nr}	{N^{\rm reg}}			
\newcommand{\Ni}	{N^{\rm iso}}			
\newcommand{\wi}	{w^{\rm iso}}			
\renewcommand{\wr}	{w^{\rm reg}}			
\newcommand{\fsa}	{f^{\rm s}}			
\newcommand{\Ns}	{N^{\rm s}}			
\newcommand{\fc}	{f^{\rm class}}			
\newcommand{\fe}	{f^{\rm iso}}			
\newcommand{\fr}	{f^{\rm reg}}			
\newcommand{\rhoo}	{\rho^{\rm obs}}		
\newcommand{\rhoi}	{\rho^{\rm iso}}		
\newcommand{\pr}	{p^{\rm reg}}			
\newcommand{\pe}	{p^{\rm iso}}			
\newcommand{\Preg}	{P^{\rm reg}}			
\newcommand{\Piso}	{P^{\rm iso}}			
\newcommand{\Qden}	{Q^{\rm den}}			
\newcommand{\Qkin}	{Q^{\rm kin}}			

\newcommand{\lambdan}	{\lambda_{\rm norm}}
\newcommand{\rhos}	{\rho_{\rm s}}			
\newcommand{\rhod}	{\rho_{\rm d}}			
\newcommand{\Ds}	{D_{\rm s}}			
\newcommand{\Dd}	{D_{\rm d}}			
\newcommand{\Mb}	{{M}_\Box}			
\newcommand{\Mo}	{{M}^{\rm obs}}		
\newcommand{\Mr}	{{M}^{\rm reg}}		

\newcommand{\ov}[1]	{\overline{#1}}			
\newcommand{\ave}[1]	{\left\langle{#1}\right\rangle}	
\newcommand{\avet}[1]	{\langle{#1}\rangle}		
\renewcommand{\b}[1]	{\bmath{#1}}			
\newcommand{\bw}	{\b{w}}
\newcommand{\bv}	{\b{v}}
\newcommand{\br}	{\b{r}}
\newcommand{\bs}	{\b{s}}
\newcommand{\bp}	{\b{p}}
\newcommand{\bJ}	{\b{J}}
\newcommand{\bt}	{\b{\theta}}

\newcommand{\bephi}	{{\hat{\b e}_{\phi}}}
\newcommand{\bel}	{{\hat{\b e}_{\ell}}}		
\newcommand{\eps}	{\epsilon}

\newcommand{\Peff}	{\Phi_{\rm eff}}

\newcommand{\vlos}	{\mbox{$v_{\rm los}$}}			
\newcommand{\siglos}	{\mbox{$\sigma_{\rm los}$}}		
\newcommand{\ml}	{\mbox{${\mu_\ell}$}}			
\newcommand{\mul}	{\mbox{${\mu_\ell}$}}			
\newcommand{\mb}	{\mbox{${\mu_b}$}}			
\newcommand{\mub}	{\mbox{${\mu_b}$}}			
\newcommand{\sigl}	{\mbox{${\sigma_\ell}$}}		
\newcommand{\sigb}	{\mbox{${\sigma_b}$}}			
\newcommand{\sigrl}	{\mbox{$\sigma^2_{{\rm los},\ell}$}}	
\newcommand{\sigrb}	{\mbox{$\sigma^2_{{\rm los},b}$}}	
\newcommand{\siglb}	{\mbox{$\sigma^2_{\ell b}$}}		

\newcommand{\msun}	{{\rm M}_\odot}
\newcommand{\kms}	{{\rm\,km\,s}^{-1}}
\newcommand{\kpc}	{{\rm\,kpc}}
\def\pc{\,{\rm pc}}
\newcommand{\masyr}	{{\rm\,mas\,yr}^{-1}}
\newcommand{\window}[2] {\mbox{($#1\degr\!,\,#2\degr$)}}	
\newcommand{\wBaa}	{\window{ 1   }{-4   }}			
\newcommand{\wMia}	{\window{ 8   }{ 7   }}			
\newcommand{\wMib}	{\window{12   }{ 3   }}			
\newcommand{\wTT}	{\window{ 8.4 }{-6   }}			
\newcommand{\wBla}	{\window{ 1.21}{-1.67}}			
\newcommand{\wBlb}	{\window{-1.14}{ 1.81}}			

\newcommand{\App}[1]	{Appendix \ref{app:#1}}			
\newcommand{\Eqn}[1]	{Equation (\ref{#1})}			
\newcommand{\eqn}[1]	{equation (\ref{#1})}			
\newcommand{\eqnsto}[2]	{equations (\ref{#1}) to (\ref{#2})}	
\newcommand{\beqn}[1]	{(equation~\ref{#1})}			
\newcommand{\tab}[1]	{Table~\ref{tab:#1}}			
\newcommand{\fig}[1]	{Fig.~\ref{fig:#1}}			
\newcommand{\figsto}[2]	{Figs.~\ref{fig:#1} to \ref{fig:#2}}	
\newcommand{\figsand}[2]{Figs.~\ref{fig:#1} and \ref{fig:#2}}	

\ifoldfss
  \ifCUPmtlplainloaded \else
    \NewTextAlphabet{textbfit} {cmbxti10} {}
    \NewTextAlphabet{textbfss} {cmssbx10} {}
    \NewMathAlphabet{mathbfit} {cmbxti10} {} 
    \NewMathAlphabet{mathbfss} {cmssbx10} {} 
  \fi
  \ifAMStwofonts
    \ifCUPmtlplainloaded \else
      \NewSymbolFont{upmath} {eurm10}
      \NewSymbolFont{AMSa} {msam10}
      \NewMathSymbol{\upi}     {0}{upmath}{19}
      \NewMathSymbol{\umu}     {0}{upmath}{16}
      \NewMathSymbol{\upartial}{0}{upmath}{40}
      \NewMathSymbol{\leqslant}{3}{AMSa}{36}
      \NewMathSymbol{\geqslant}{3}{AMSa}{3E}

       \let\le=\leqslant
       \let\ge=\geqslant
    \fi
  \fi
\fi 

\ifnfssone
  \newmathalphabet{\mathit}
  \addtoversion{normal}{\mathit}{cmr}{m}{it}
  \addtoversion{bold}{\mathit}{cmr}{bx}{it}
  \newmathalphabet{\mathbfit} 
  \addtoversion{normal}{\mathbfit}{cmr}{bx}{it}
  \addtoversion{bold}{\mathbfit}{cmr}{bx}{it}
  \newmathalphabet{\mathbfss} 
  \addtoversion{normal}{\mathbfss}{cmss}{bx}{n}
  \addtoversion{bold}{\mathbfss}{cmss}{bx}{n}
  \ifAMStwofonts
    \ifCUPmtlplainloaded \else
      \UseAMStwoboldmath
      \makeatletter
      \new@mathgroup\upmath@group
      \define@mathgroup\mv@normal\upmath@group{eur}{m}{n}
      \define@mathgroup\mv@bold\upmath@group{eur}{b}{n}
      \edef\UPM{\hexnumber\upmath@group}
      \new@mathgroup\amsa@group
      \define@mathgroup\mv@normal\amsa@group{msa}{m}{n}
      \define@mathgroup\mv@bold\amsa@group{msa}{m}{n}
      \edef\AMSa{\hexnumber\amsa@group}
      \makeatother
      \mathchardef\upi="0\UPM19
      \mathchardef\umu="0\UPM16
      \mathchardef\upartial="0\UPM40
      \mathchardef\leqslant="3\AMSa36
      \mathchardef\geqslant="3\AMSa3E

       \let\le=\leqslant
       \let\ge=\geqslant
    \fi
  \fi
\fi 

\ifnfsstwo
  \DeclareMathAlphabet{\mathbfit}{OT1}{cmr}{bx}{it}
  \SetMathAlphabet\mathbfit{bold}{OT1}{cmr}{bx}{it}
  \DeclareMathAlphabet{\mathbfss}{OT1}{cmss}{bx}{n}
  \SetMathAlphabet\mathbfss{bold}{OT1}{cmss}{bx}{n}
  \ifAMStwofonts
    \ifCUPmtlplainloaded \else
      \DeclareSymbolFont{UPM}{U}{eur}{m}{n}
      \SetSymbolFont{UPM}{bold}{U}{eur}{b}{n}
      \DeclareSymbolFont{AMSa}{U}{msa}{m}{n}
      \DeclareMathSymbol{\upi}{0}{UPM}{"19}
      \DeclareMathSymbol{\umu}{0}{UPM}{"16}
      \DeclareMathSymbol{\upartial}{0}{UPM}{"40}
      \DeclareMathSymbol{\leqslant}{3}{AMSa}{"36}
      \DeclareMathSymbol{\geqslant}{3}{AMSa}{"3E}

       \let\le=\leqslant
       \let\ge=\geqslant
    \fi
  \fi
\fi 

\ifCUPmtlplainloaded \else
  \ifAMStwofonts \else 
    \def\upi{\pi}
    \def\umu{\mu}
    \def\upartial{\partial}
  \fi
\fi

 \def\figdir {.}		

\title{A Dynamical Model of the Inner Galaxy}
\author[Ralf M.\ H\"afner, N. Wyn Evans, Walter Dehnen and James Binney]
       {Ralf H\"afner, N. Wyn Evans, Walter Dehnen and James Binney\\
        Theoretical Physics, Department of Physics, 1 Keble Rd, 
        Oxford, OX1 3NP}

\begin{document}

\maketitle

\label{firstpage}

\begin{abstract}
An extension of Schwarzschild's (1979) galaxy-building technique is presented
that, for the first time, enables one to build Schwarzschild models with
known distribution functions (DFs). The new extension makes it
possible to combine a DF that depends only on classical integrals with orbits
that respect non-classical integrals. With such a
combination Schwarzschild's orbits are used only to represent the difference
between the true galaxy DF and an approximating classical DF.

The new method is used to construct a dynamical model of the inner Galaxy.
The model is based on an orbit library that contains $22\,168$ regular
orbits. The model aims to reproduce the three-dimensional mass density of
Binney, Gerhard \& Spergel (1997), which was obtained through deprojection
of the COBE surface photometry, and to reproduce the observed kinematics in
three windows -- namely Baade's Window with $(\ell, b)\eq\wBaa$ and two
off-axis fields at \wMia\ and \wMib. The viewing angle is assumed to be
$20\degr$ to the long axis of the bar and the pattern speed is taken to be
$60\kms\kpc^{-1}$.

The model fits essentially all the available data within the
innermost 3\kpc. The axis ratio and the morphology of the projected density
contours of the COBE bar are recovered to excellent accuracy within corotation.
The kinematic quantities -- the line-of-sight streaming velocity and velocity
dispersion, as well as the proper motions when available -- are recovered, not
merely for the fitted fields at \wBaa\ and \wMia, but also for three new fields
at \wTT, \wBla, and \wBlb. The dynamical model deviates most from the input 
density close to the Galactic plane just outside corotation, where the 
deprojection of the surface photometry is suspect. The dynamical model does not
reproduce the kinematics at the most distant window, \wMib, where disk 
contamination of the data may be severe.

Maps of microlensing optical depth are presented both for randomly chosen
stars and for stars that belong to individual components within the model.
While the optical depth to a randomly chosen star in Baade's Window is half
what measurements imply, the optical depth to stars in a particular
component can be as high as the measured values.
The contributions to the optical depth towards randomly chosen stars from
lenses in different components are also given.
 \end{abstract}


\section{INTRODUCTION}
 We live rather far out in the disc of a spiral galaxy, so studies of the solar
neighbourhood do not provide a balanced view of the Galaxy as a whole. It is
essential to complement these studies with investigations of the inner Galaxy.
This more remote region is hard to study because it is almost completely
obscured by dust at optical wavelengths. Consequently, the first indication that
ours is a barred galaxy came from 21-cm observations (de Vaucouleurs 1964). It
is only in the last few years that a combination of radio-frequency (Binney et
al.\ 1991) and near infrared studies (Blitz \& Spergel 1991) have demonstrated
beyond reasonable doubt that the inner Galaxy is dominated by a bar whose nearer
end lies at positive longitudes. Evidence from microlensing has also been
adduced in favour of a barred inner Galaxy (e.g., Paczy\'nski et al.\ 1994,
Evans 1994), although this remains a  controversial matter (e.g.,
Bissantz et al.\ 1997, Sevenster et al.\ 1998).

A promising way to constrain the characteristics of the bar is to combine
observations at various wavelengths with theory by trying to construct a {\it
dynamical model} of the inner Galaxy that is compatible with the available data.
This is the only true test of our assumptions regarding the Galaxy.  Of course,
the current observational data are too scanty to allow us to infer a unique
model, and, unsurprisingly, there is controversy as to the viewing angle, length
and pattern speed of the Galactic bar (e.g., Binney et al.\ 1991, Blitz \&
Spergel 1991, Sevenster et al.\ 1998).  The real value of a dynamical model is
that it gives us predictive power and suggests further observational tests to
confirm or constrain the structural parameters of the bar. Pioneering studies
along these lines include those of Pfenniger \& Friedli (1991), Sellwood (1993),
Zhao (1996) and Sevenster et al.\ (1998).
 
In this paper we build a dynamical model by populating orbits in a given
potential. This technique was pioneered by Schwarzschild (1979, 1982, 1993) and
further developed by Merritt \& Fridman (1996) and Zhao (1996). Schwarzschild's
original application of his method was to the building of models of elliptical
galaxies in which most of phase space was regular. In general, rotating barred
potentials support very considerable numbers of irregular orbits and a
straightforward application of Schwarzschild's method is not profitable.
In Section 2 we therefore extend Schwarzschild's method in such a way that both regular
and chaotic
regions of phase space can be populated.

Our model of the inner Galaxy is constrained to fit the three dimensional
luminosity density of Binney, Gerhard \& Spergel (1997), which was designed
to reproduce the infrared data from the COBE satellite.  It was recovered by
Richardson--Lucy deconvolution of the COBE data, after correction for
extinction using a model in which dust is distributed throughout the Galaxy
(Spergel, Malhotra \& Blitz 1996). The resulting three dimensional density
is specified on $59\times59\times39$ data-cube, corresponding to a box that
is $10\kpc$ on a side and $2.8\kpc$ thick. Our dynamical model is also
constrained to reproduce the kinematic observations in three windows --
Baade's Window with $(\ell,b)\eq\wBaa$ and the two off-axis fields of
Minniti et al.\ (1992) at \wMia\ and \wMib. All our constraints are
described in Section 3, where particular attention is paid to the best ways
of comparing models to kinematic data. To have value, a dynamical model
requires a large number of orbits, which are the basic building blocks in
Schwarzschild's method. Our orbit library, which contains the density and
kinematic contributions of over $23\,000$ orbits, is described in Section 4.
Section 5 discusses the penalty and merit functions used to drive the
mass on the orbits towards the desired input density. Our final dynamical
model of the inner Galaxy is analyzed in detail in Section 6. There we
predict the values of measurable kinematic quantities in several fields near
the Galactic centre, describe the model's DF, and analyze the variation of
optical depth to gravitational microlensing both with position on the sky
and with the Galactic component to which either the source star or the
lensing object belongs. Section 7 sums up.

\section{EXTENDING SCHWARZSCHILD'S TECHNIQUE}
 \subsection{Splitting the DF}
 In Schwarzschild's original technique, we calculate $N$ orbits labelled by
$i\eq1,\ldots,N$ in the given potential and determine the fraction $p_{ij}$ of
the time that the $i$th orbit lies in each of the $j\eq1,\dots,K$ cells. Let
$\rhoo_j$ be the system's original density in the $j$th cell with volume $V_j$.
Then we seek the non-negative weights $w_i$ that minimize the discrepancies
 \beq\label{defsDeltaj}
	\Delta_j = V_j \rhoo_j - 
		\frac{M}{N}\,\sum\limits_{i=1}^{N}
		w_i\; p_{ij}.
\eeq
 Here, the constant factors can be absorbed into the definition of the weights,
but we have written them explicitly for later convenience.

So long as the galaxy model is stationary in inertial space or in a frame of
reference rotating at constant pattern speed, the Hamiltonian $H$ is an
isolating integral of the equations of stellar motion. If the system is
axisymmetric, one component of angular momentum, $L_z$, will also be an
isolating integral. Consequently, by Jeans'
theorem, any DF that is a function of $H$, and where appropriate of $L_z$,
will satisfy the collisionless
Boltzmann equation (see e.g., Binney \& Tremaine 1987). This suggests that we
approximate the true DF by a classical  DF,  $\fc(H)$ or $\fc(H,L_z)$ as
appropriate.  Since $H$ and $L_z$ are known
functions of the conventional phase-space coordinates
 \beq
	\bw\equiv(\br,\bp),
\eeq
we can readily calculate the value of the density or any kinematic
quantity to which $\fc$ would give rise. Except in special cases, the true DF,
$f$, will depend on non-classical isolating integrals. Therefore we write
 \beq
	f(\bw)=\fc\big(H(\bw),L_z(\bw)\big)+\fr(\bw),
\eeq
where $\fr$ is a function that depends on non-classical isolating integrals.
We show below how Schwarzschild's technique can be used to determine
simultaneously  $\fc$ and $\fr$ under the assumption that
non-classical isolating integrals exist only along orbits that are regular
in the sense that they have three effective isolating integrals. The latter
we take to be actions $J_i,\,i\eq1,2,3$.  For simplicity
of exposition we henceforth assume that the only classical integral is $H$,
which is certainly the case for the inner Galaxy. Since DFs that depend on
$H$ alone generate isotropic velocity distributions in the appropriate frame
of reference, we refer to the component generated by $\fc$ as the {\em
isotropic component\/} and call its DF the isotropic DF, $\fe=\fc(H)$. All
the results below generalize trivially to more general classical DFs.

Note that in our convention
$f(\bw)$ is a probability density and is always normalized to unit phase-space
integral inside the box used to fit the density (see below).

\subsection{From weights to DF}
 The key problem involved in representing $\fr\eq f\mi\fe$ with Schwarzschild's
technique, is the determination of the value of the DF that is implied by a
given set of weights $\wr_i$. We solve this problem by using an arbitrary,
everywhere positive, correctly normalized probability density $\fsa(\bw)$ to
sample points in phase-space. With these points as initial conditions, we
integrate along orbits. Then we define
 \beq
	\ov\fsa\big(\bJ(\bw)\big)=\bec
	\ds \!\lim_{T\to\infty} {1\over T} \int_0^T\!\d t\,
		\fsa\big(\tilde{\bw}(t)\big)		&
		$\bea{l} \mbox{if the orbit} \\[-.2ex] 
			\mbox{is regular,} \eea $		\\[.5ex]
	0	&			$\,$otherwise,
\eec \eeq
 where $\tilde{\bw}(t)$ is the orbit with initial conditions ${\bw}$.
Constructed thus, $\ov\fsa(\bJ)$ automatically obeys the collisionless Boltzmann
equation.  Hence, any distribution function consisting of regular orbits only
can be written as
 \beq\label{magic}
	\fr(\bJ)=\wr(\bJ)\,\ov{\fsa}(\bJ),
\eeq
 where $\wr(\bJ)$ is some weight-function. Moreover, $\ov\fsa(\bJ)$ gives the
relative probability that a regular orbit obtained by sampling phase space
according to the probability density $\fsa(\bw)$ will be the orbit with actions
$\bJ$. The corresponding probability-density for picking the $i$th regular
orbit, is $(2\upi)^3(\Ns/\Nr)\ov\fsa (\bJ_i)$, where $\Ns$ initial conditions 
gave rise to $\Nr$ regular orbits, while the factor of $(2\upi)^3$ accounts for
the phase-space volume at constant $\bJ$.

The isotropic part of the DF, $\fe$, is most conveniently represented by a
superposition of basis functions $B_i(\EJ)$, for example splines, \beq
\label{fiso}
	\fe(\bw) = \sum_i \wi_i\,B_i\big(H(\bw)\big), 
\eeq
such that the complete DF is determined by the set of weights 
$\{\wi_i,\wr(\bJ)\}$.

\subsection{The non-negativity constraint}
 In irregular regions of phase space, $\fr$ vanishes and the total DF, $f$, is
equal to the isotropic DF, $\fe$. Hence the latter must be non-negative in
irregular regions. By \eqn{magic}, the requirement that the total DF be
non-negative translates to the following constraint on the weights:
 \beq\label{truecon}
	\wr(\bJ)\ge-\frac{\fe\big(H(\bJ)\big)}{\ov\fsa(\bJ)}
	=-\frac{\sum_i \wi_i\,B_i\left(H(\bJ)\right)}{\ov\fsa(\bJ)}
\eeq
 In general, $\fe$ will be everywhere non-negative and this equation allows
$\wr(\bJ)$ a limited degree of negativity.

The physical implication of an orbit having a negative weight is this.
Frequently, both regular and irregular orbits exist at a given value of $H$.
In a three-dimensional system, irregular orbits tend to Arnold diffuse over
much of the phase-space hypersurface of constant $H$. They are, however,
rigorously excluded from regions of the hypersurface that are occupied by
regular orbits.  The isotropic part of the DF may assign non-zero density to
irregular orbits at some value of $H$ when the total phase-space density,
$f$, in some regular region of the same energy is negligible. In such a case
the weights of the regular orbits would approximately satisfy the equality
condition in \eqn{truecon}. Hence, regular orbits can be important even if
no stars are on them by virtue of their ability to exclude stars on
irregular orbits from subsets of their energy hypersurfaces.

\subsection{From observations to weights} \label{sec:formal}
 To calculate $f$, the weights $\{\wi_i,\wr(\bJ)\}$ must be determined from
observational data. Any observable moment, such as the density, is linear in the
DF, and can be split into the contributions arising from $\fe$ and $\fr$. That
is
 \beq \label{moment}
	\Pi[f] = \int\d^6\bw\, f(\bw)\,\widehat\Pi(\bw) = \Pi[\fe] + \Pi[\fr],
\eeq
 where $\widehat{\Pi}(\bw)$ is the function that characterizes the moment in
question -- see below. The left-hand side of \eqn{moment} is given by
observations and the right-hand side is a linear function of the weights. A
sufficiently large set of such equations, for different functions $\widehat\Pi$, can
now be treated as an inverse problem and a standard technique used to recover
the weights $\{\wi_i,\,\wr(\bJ)\}$ from the measured values of the left-hand
sides.

Since $H(\bw)$ is a known function, the moments $\Pi[\fe]$ are straightforward
to evaluate (see Section~\ref{sec:fiso}). Using (\ref{magic}) and the standard
identity $\d^6w\id\d^3\!\bJ\,\d^3\bt$, where the $\theta_i$ are the angle
variables conjugate to the $J_i$, the moments arising from $\fr$ can be written
as
 \beq\label{Pioff}
	\Pi[\fr] =	\int\d^3\!\bJ\;\wr(\bJ)\;
			\ov{\fsa}(\bJ)\int\d^3{\bt}\;\widehat{\Pi}
			\big(\bw(\bJ,{\bt})\big).
\eeq
 The right part of the right-hand side can be calculated numerically via the
time-averaging theorem (e.g., Binney \& Tremaine 1987):
 \beq\label{timeave}
	\ov\Pi(\bJ)\equiv(2\upi)^{-3}\!\int\d^3{\bt}\;\widehat{\Pi}(\bw) =
	\lim_{T\to\infty}\frac{1}{T}\int_0^T\d t\;
	\widehat{\Pi}\big(\bw(t)\big).
\eeq
 The integral over $\d^3\!\bJ$ in \eqn{Pioff} is performed by Monte-Carlo
integration employing $\Nr$ regular orbits sampled from the probability density
to $(2\upi)^3(\Ns/\Nr)\fsa(\bJ)$ that was computed above. This yields:
 \beq \label{Pisum}
	\Pi[\fr] = 	\frac{1}{\Ns}\,\sum_{i=1}^{\Nr}
			\wr_i\;\ov{\Pi}(\bJ_i),
\eeq
 where $\wr_i$ is the weight of the $i$th orbit and $\ov\Pi_i$ is the time 
average of $\widehat\Pi(\bw)$ on this orbit. Note that in practice we do not know 
$\wr(\bJ)$ but $\wr_i$, and hence $\wr \big(\tilde{\bw}_i(t)\big)$, for the 
sampled orbits only.

As an example of this formalism, consider $\Mr_j$, the contribution from 
regular orbits to the mass in cell $j$. In this case we define
 \beq
	\widehat\Pi_j(\bw)=\bec
		1 & 	if $\bw$ lies in cell $j$,	\\[0.2ex]
		0 &	otherwise.
\eec \eeq
 Then the phase-space integral over $\widehat{\Pi}_i$, i.e.\ the left-hand side 
of \eqn{Pioff}, becomes the fraction of the total mass that is in cell $j$. 
Here, ``total mass'' refers only to the mass $\Mb$ inside the box that contains 
all cells -- recall that $f(\bw)$ is normalized to unit phase-space integral 
inside that box. Hence, multiplying \eqn{Pisum} by $\Mb$ we have the mass 
$\Mr_j$ on regular orbits in cell $j$
 \beq\label{rhofromrho}
	\Mr_j= \frac{\Mb}{\Ns}\sum_{i=0}^{\Nr}\wr_i\;\pr_{ij},
\eeq 
where 
\beq \label{pij}
	\pr_{ij} = \lim_{T\to\infty}\frac{1}{T}\int_0^T\d t\;
			\widehat{\Pi}_j\big(\bw_i(t)\big)
\eeq
 is the fraction of the time that the $i$th regular orbit is in the $j$th cell.
The analogy between this equation and \eqn{defsDeltaj} is clear.

\section{APPLICATION TO THE GALAXY}
 We will be working in a rotating frame of reference, in which the dynamics are
governed by the Hamiltonian
 \beq\label{defham}
	H(\bw) = \half{\bp}^2+\Phi(\br)-\b{\Omega}\cdot(\br \times\bp),
\eeq
 where $\bp$ is the canonical momentum per unit mass and $\b{\Omega}$ is the
angular velocity of the bar. Standard manipulations enable us to recast this
into the form (e.g., Binney \& Tremaine 1987)
 \beq\label{defpeff}
	H(\bw) = \half\bv^2+\Peff(\br),\quad
	\Peff(\br)=\Phi(\br)-\half\Omega^2(x^2\pl y^2)
\eeq
 where $\bv$ is the velocity in the rotating frame and $\Peff$ is the effective
potential. The Hamiltonian $H$, which is an exactly conserved quantity, is also
known as Jacobi's integral $\EJ$.

\begin{figure}
\epsfxsize\hsize\epsfbox[93 280 521 772]{\figdir/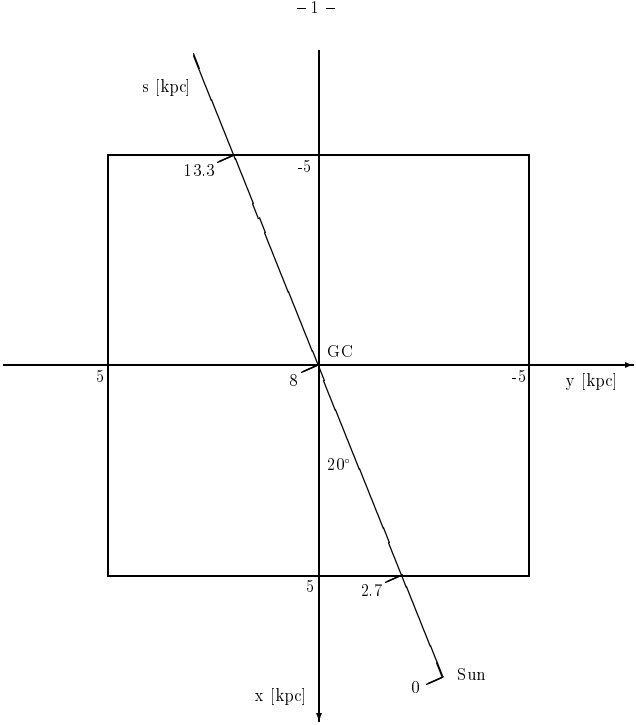}
\caption[]{This shows the geometry of the bar in the Galactic plane. The three 
	dimensional luminosity density of Binney et al.\ (1997) resides in a box
	that is $5\kpc$ square in the plane. The viewing angle of the Sun is 
	taken as $20\degr$. The $x$-axis corresponds to the bar's major axis, 
	the $y$-axis is the minor. The intercepts of the line of sight through 
	the Galactic Centre with the box are marked. The bar is rotating in the 
	clockwise direction. \label{fig:geom}}
\end{figure}

\subsection{The density}
 As the original density $\rho^{\rm o}$, we use the three-dimensional model that
Binney et al.\ (1997) obtained by deprojecting the near-infrared COBE surface
photometry.  The surface photometry employed by Binney et al.\ had been
corrected for absorption by dust by a procedure that is outlined in Spergel,
Malhotra \& Blitz (1996). The non-parametric Richardson--Lucy algorithm used by
Binney et al.\ is based on the assumption that the density is eight-fold
symmetric (with respect to the major, intermediate and minor axes). While this
assumption is reasonable enough for the bar itself, it causes features like
spiral arms to be incorrectly reproduced -- see Binney et al.\ for a discussion
of this problem.  Both the viewing angle and the pattern speed of the Galactic
bar are somewhat controversial (e.g., Binney et al.\ 1991, Sevenster et al.\
1998). For the sake of definiteness, we take the bar's viewing angle as
$20\degr$ and the pattern speed as $60 \kms\kpc^{-1}$.  The Sun is assumed to
lie at a Galactocentric radius of $8\kpc$ and $0.014\kpc$ above the Galactic
plane.  The circular speed at $R_0$ is taken to be $200\kms$. The Sun's peculiar
motion is $(10,5,7)\kms$, where the first component is along the line of sight
towards the Galactic centre, the second is in the direction of Galactic rotation
and the third component points towards the north Galactic pole. In other words, the
Sun is moving in towards the Galactic centre, leads the local standard of rest
and moves up and away from the Galactic plane (see e.g., Binney \&
Merrifield 1998). All this means that in the rest frame of the Galaxy with
($x,y,z$) coordinates aligned with the symmetry axes of the bar, the Sun has
phase-space coordinates
 \beq \bea{rcl}
	(x,y,z) 	&=& 	(7.5,2.7,0.014)  \kpc, \\[0.5ex]
	(p_x,p_y,p_z) 	&=&	(60.7,-196, 6.98)\kms .
\eea \eeq
 The other coordinate system we frequently use is a heliocentric frame. The
line-of-sight distance $s$ and Galactic longitude and latitude ($\ell,b$) are
the configuration-space coordinates. Velocity space is given by radial or
line-of-sight velocity \vlos, together with the proper motions $(\mul,\mub)$. In
this system, the Galactic centre has coordinates
 \beq \bea{rcl}
	(s,\ell,b) 	&=& (8,\, 0,\, 0) \kpc, \\[0.2ex]
	\vlos 		&=& -10\kms,		\\[0.2ex]
	(\ml,\mb) 	&=& (-5.4,-0.18)\masyr
\eea \eeq
 Note that at the Galactic centre, $1\masyr\eq37.92\kms$.

\begin{figure*}
\epsfxsize=\hsize \epsfbox[80 560 510 775]{\figdir/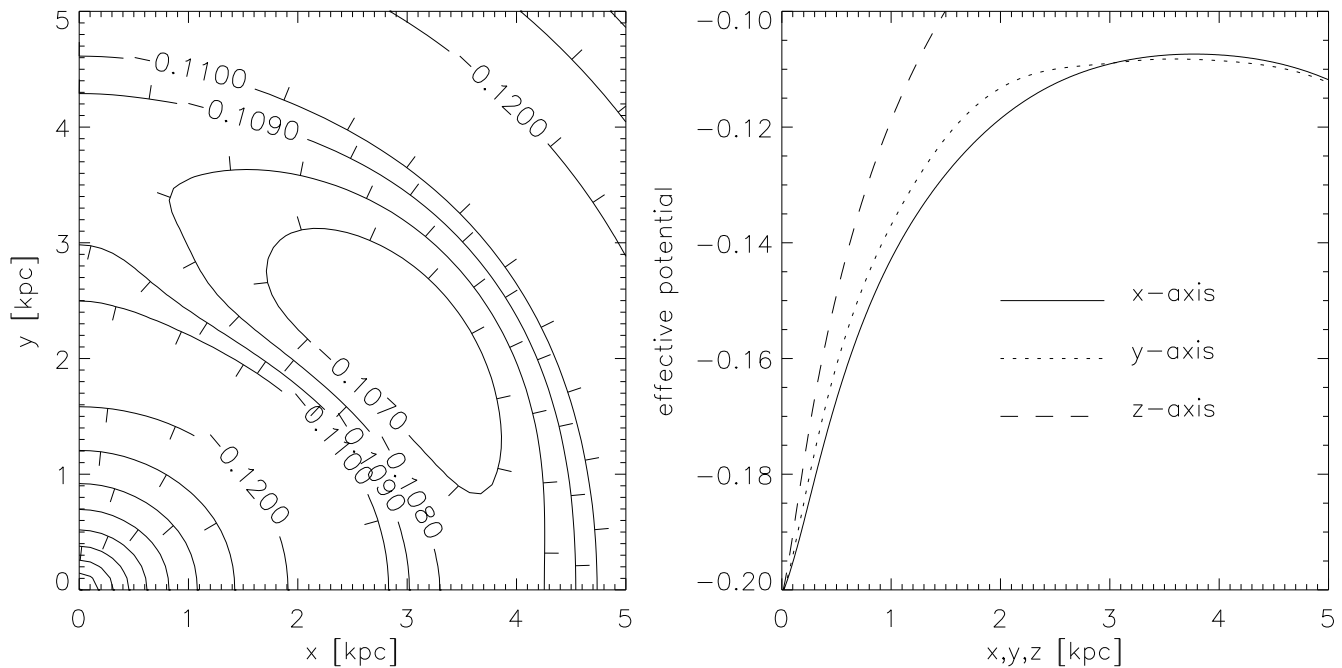}
\caption[]{The left panel shows contours of the effective potential in the 
	Galactic plane. The right panel shows the effective potential along 
	the major, intermediate and minor axes. The minimum of the effective 
	potential is at the Galactic Centre, while maxima occur both on the 
	$y$-axis and at $(x,y)\eq({\pm}2.9,\,{\pm}2.2)\kpc$. There are saddle
	points on the $x$-axis and close to the maxima along the $y$-axis.
	\label{fig:potfig}}
\end{figure*}

The density plays a dual role in Schwarzschild's method: it both constrains the
weights through \eqn{defstriang} and, through Poisson's equation, specifies the
potential in which the orbits are calculated. The Galactic density of Binney et
al.\ (1997) comes in two parts.  First, inside a box that is $10\kpc$ on a side
and $2.8\kpc$ thick and whose geometry and position w.r.t.\ the Sun is shown in
\fig{geom}, the density is specified on $59\nc{\times}59\nc{\times}39$ grid
points.  This density distribution was obtained by non-parametric
Richardson--Lucy deprojection of the photometry, followed by multiplication by a
constant mass to light ratio $\Upsilon$.  Between grid points, the density is
evaluated through three-dimensional cubic splines.  Outside the box the density
is given by an analytic function that is based on the work of Spergel, Malhotra
\& Blitz (1996):
 \beq\label{firste} 
	\rho(\br)=\rho_0\big[f_{\rm b}(\br)+f_{\rm d}(\br) \big] 
\eeq
 where
\bsq \ben
	f_{\rm b} &\equiv& 
		f_0{\e^{-a^2/a_{\rm m}^2}\over(1+a/a_0)^{1.8}},	\\
	f_{\rm d} &\equiv&
		\bigg({\e^{-|z|/z_0}\over z_0}
		+\alpha{\e^{-|z|/z_1}\over z_1}\bigg)
		R_{\rm d}\, \e^{-R/R_{\rm d}},			\\
	a	  &\equiv&
		\bigg(x^2\pl{y^2\over\eta^2}\pl{z^2\over\zeta^2}
		\bigg)^{\!1/2}\;\;\hbox{and}\quad
		R\equiv\sqrt{x^2\pl y^2}.
\een \esq
 The constants in these equations are as follows: $\rho_0\eq2.05\nc{
\times}10^8\msun\kpc^{-3}$ for the $L$-band, $f_0\eq624$, $a_{\rm m}\eq
1.9\kpc$, $a_0\eq100\pc$, $R_{\rm d}\eq2.5\kpc$, $z_0\eq210\pc$, $z_1\eq42\pc$,
$\alpha\eq0.27$, $\eta\eq0.5$, and $\zeta\eq0.6$.  Physically, these equations
specify an exponential disk, in which the vertical distribution is the sum of
two exponentials, and a triaxial bulge that extends to $R\nc{\sim}2\kpc$. Since
we use the density distribution only at $R\nc{>}5\kpc$, the bulge has a
negligible impact on our model. 
box and the density exterior to the box is achieved by setting $\rho\eq
g\rho_{\rm box}+ (1\mi g)\rho_{\rm ext}$, where
 \beq
	g \equiv \exp\Big[-{s^8 \over1-s^8}\Big]
	\qquad\qquad 0\le s\le 1
\eeq
 and $s^4\id(x/5\kpc)^4\pl (y/5\kpc)^4\pl (z/1.4\kpc)^4$. Appendix A describes
the technique used to determine the gravitational potential $\Phi$ generated by
this mass distribution.  The interesting features of the potential are
conveniently described in the rotating $(\br,\bv\id\dot{\br})$ frame of the bar.
Contours of the effective potential~(\ref{defpeff}) are plotted in \fig{potfig}.
The plot is dominated by a basin whose rim lies at $R\nc{\sim}3.8\kpc$. The
lowest point on the rim defines the critical value of the potential, $\Peff^{\rm
rim}$. Orbits whose Jacobi energy $E_{\rm J}$ is smaller $\Peff^{\rm rim}$
cannot cross the rim. Consequently, orbits belonging to the isotropic component
with $E_{\rm J}\nc{<}\Peff^{\rm rim}$ are of two types: those that lie entirely
inside the rim, and those that lie outside it. The latter extend to infinity, as
do orbits with $E_{\rm J}\nc{\ge}\Peff^{\rm rim}$. Since such unbound orbits are
useless for galaxy modelling, $\fe$ is non-zero only for $E_{\rm
J}\nc{<}\Peff^{\rm rim}$, and then describes orbits that lie entirely within the
rim.

In \fig{potfig} a maximum centred on $(2.9,2.2)\kpc$ is conspicuous. The
potential's other stationary points are the Lagrange points ${\rm L_1}$ to $\rm
L_5$ that were first identified in the context of the restricted three body
problem.  $\rm L_{1,2,3}$ lie along the $x$-axis, with $\rm L_1$ being at the
Galactic Centre and $\rm L_{2,3}$ opposite each other at $x\ap\pm3.7\kpc$, while
$\rm L_{4,5}$ lie on the $y$-axis at $y\ap\pm 3.5\kpc$.

\subsection{The kinematics}
\subsubsection{The selection function}
 The main problem when observing stars in the bulge is obscuration. This problem
is worst within the Galactic plane, and a relatively unobscured optical view at
the Galactic centre is only possible in a very few windows, the most famous of
which is Baade's Window at $(\ell,b)\eq\wBaa$.

To compare observational data with a dynamical model, careful thought has to be
given to the selection criteria applied to obtain the sample of observed stars. 
The selection function $\eps(s,M)$ is the probability that a star of absolute
magnitude $M$ that lies at heliocentric distance $s$ is included in the sample.
Unhappily, for many published surveys $\eps(s,M)$ is hard to determine.  If a
survey contains all stars brighter than the limiting magnitude $m_{\rm max}$ and
fainter than some cutoff magnitude some magnitude $m_{\rm min}$, the function
$\eps(M,s)$ is given by
 \beq\label{eff_ms}
	\eps(M,s) = \bec
		1&:$\; m_{\rm min}<M\pl5\log{\ds s\over\ds10{\rm pc}}
			\pl\gamma s<m_{\rm max}$\\
		0&: otherwise
\eec \eeq
 where $\gamma$ is the differential extinction in magnitudes per unit distance.
Direct observations of stars and their colour excess yield the total extinction
$A$ in certain windows, but $\gamma$ itself is only available from
three-dimensional dust-models whose reliability is uncertain.  Even if a dust
model fits the dust distribution well when the latter is averaged over some
scale, it will be inaccurate at individual points because extinction within the
disk is very patchy. Therefore, in \eqn{eff_ms} we use the measured total
extinction $A$ rather than the differential extinction $\gamma$ of a model --
since most extinction lies in a foreground screen close to the Sun, this
procedure should not give rise to large errors. To allow for the patchiness of
extinction, we treat $A$ as a Gaussian random variable.

If the $\phi(M)$ is the stellar luminosity function,  then
the general selection function $\epsilon(s)$
 \beq 
	\eps(s) = \int\d M \, \phi(M) \; \eps(M,s), 
\eeq 
 gives the probability that a star at distance $s$ that has unknown absolute
magnitude will included in a survey. Combining the last two equations, we obtain
the family of generic selection functions that we will use:
 \beq \label{generic} \bea{rcl}
	\eps(s) & \ds = \frac{1}{\sqrt{2\upi\sigma^2_A}} & \ds \int
         \d A\, \e^{-A^2/2\sigma_A^2} \\
        && \ds \times \int\limits
		^{m_{\rm max}-A-5\log[s/10{\rm pc}]}
		_{m_{\rm min}-A-5\log[s/10{\rm pc}]}
		\d M\;\phi(M). \eea\eeq
 The samples in the bulge are normally dominated by giant stars.  Each type of
star (e.g., K giants or M giants) may be considered to have a relatively narrow
band of intrinsic luminosities $M$ that is well-modelled by a Gaussian
 \beq
	\phi(M) = \frac{1}{\sqrt{2\upi\sigma^2_M}}\,
	\exp\left[-\frac{(M-M_{\rm int})^2}{2\sigma_M^2}\right].
\eeq
 Here, the average intrinsic luminosity $M_{\rm int}$ and dispersion
$\sigma_M$ vary according to stellar type.

\subsubsection{The fitted windows}
 The kinematic data are fitted in three windows. These are Baade's Window and
the two off-axis windows studied by Minniti et al.\ (1992) at $(\ell,b)\eq\wMia$
and \wMib. These windows are chosen because the selection criteria seem 
reasonably clear-cut and reproducible.

Baade's Window has been studied extensively in the visible wave-band and it is
the only one for which proper-motion data are available. Sharples, Walker \&
Cropper (1990) measured line-of-sight velocities for an unbiased sample of 239
late-type M giants. The stars were divided into two groups which show different
kinematics. The first group contains 14 bright stars ($I\nc{\le}11.8$) with a
relatively small velocity dispersion of $71^{+20}_{-11}\kms$. They are believed
to be either in the outer part of the bulge on the solar side, or foreground
disk giants, or younger, more massive asymptotic giant-branch stars. The second
group, of fainter stars, is attributed to the bulge itself. Sharples et al.\
argue that this group can be considered complete. The velocity dispersion is
considerably higher at $113^{+6}_{-5}\kms$. The mean velocity is
$4\nc{\pm}8\kms$. The value of $113^{+6}_{-5}\kms$ for the line-of-sight
velocity dispersion seems to be very robust as other studies find similar values
(Rich 1988). This dataset is modelled with one of the generic selection
functions introduced in \eqn{generic} with parameters $I_{\rm max}\eq13.4$ and
$I_{\rm min}\eq11.8$. The mean extinction $A_I\eq0.87$ and its dispersion
$\sigma_A\eq0.1$ are estimated from Table 3.21 in Binney \& Merrifield (1988).

\begin{figure}
	\epsfxsize\hsize\epsfbox[55 175 588 702]{\figdir/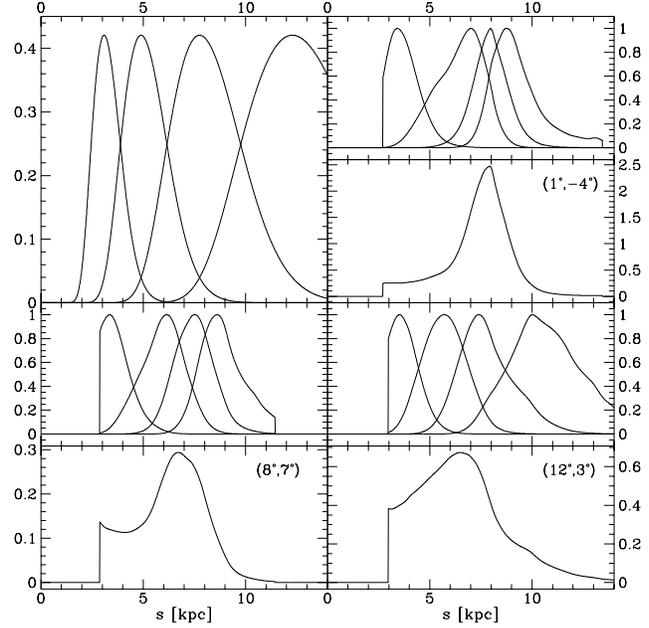}
	\caption[]{{\bf Top left:} A set of selection functions defined by
	$m_{\rm min}\eq m_{\rm max}\mi1,\; M_{\rm int}\eq0,\;A\eq0,\;\sigma_M
	\eq0.5$ and $\sigma_A\eq0.1$. $m_{\rm max}$ is taken to be 13, 14, 15 
	and 16, respectively. {\bf Top right, bottom left, bottom right:} Two 
	panels are shown for each of the windows at \wBaa, \wMia, and \wMib,
	respectively. The upper panel shows the probability of finding a star 
	at distance $s$ in a given sample while the panel on lower panel shows 
	the density along the line of sight. The sharp cut-off occurs when 
	the edge of the box is reached. \label{fig:defex}}
\end{figure}

\begin{table}
\caption[]{The parameters used in the section functions for the data-sets 
	provided by Sharples et al.\ (1990), Spaenhauer et al.\ (1992) and 
	Minniti et al.\ (1992). \label{tab:select}}
\begin{center}
\begin{tabular}{@{}r@{,$\,$}r@{$\;\;$\vline$\;\;$}
		r@{}l r@{}l r@{}l@{$\;\;$}r@{}l r@{}l@{$\;\;$}r@{}l l@{}}
\multicolumn{2}{c}{$(\ell,b)$} &
\multicolumn{2}{l}{$m_{\rm max}$} &
\multicolumn{2}{c}{$m_{\rm min}$} &
\multicolumn{4}{c}{$A\;\;\sigma_A$} &
\multicolumn{2}{c}{$m_{\rm int}$} &
\multicolumn{2}{c}{$\sigma_M$} & Ref	\\ \hline
 (1\degr&--4\degr)& 13&.4 &11&.8& 0&.87& 0&.1& --2&.75& 1&.0& Sh \\
 (1\degr&--4\degr)& 17&.5 &16&  & 1&.8 & 0&.2&   1&.3 & 1&.0& Sp \\
 (8\degr&  7\degr)& 16&.5 &12&  & 1&.5 & 0&.5& --1&.5 & 1&.0& M  \\
(12\degr&  3\degr)& 16&.5 &12&  & 1&.5 & 0&.5& --1&.5 & 1&.0& M  \\ \hline
\end{tabular}
\end{center}
The first column gives the Galactic coordinates of the window. The next two
columns report the faint $m_{\rm max}$ and bright $m_{\rm min}$ cut-offs used in
the selection function. The fourth and fifth columns give the mean extinction
$A$ and its dispersion $\sigma_A$. The mean intrinsic magnitude $M_{\rm int}$
and its dispersion $\sigma_M$ for the luminosity function of the sample are
reported in columns six and seven. These refer to the wave-bands in which the
observations are taken. The last column is the mnemonic for the window: Sh for
Sharples et al., Sp for Spaenhauer et al.\ and M for Minniti et al.
\end{table}

Spaenhauer, Jones \& Whitford (1992) conducted the first, and so far only,
survey of stars in Baade's Window that measured proper motions rather than just
line-of-sight velocities. They compared plates taken at epochs that are $~20$
years apart and selected 800 stars with $B\mi V\nc{>}1.4$. These are almost
entirely K and M giants. Out of these 800 stars, 371 were excluded due to
overcrowding or because they were too faint. Stars between $B\eq17$ and $B\eq19$
(roughly) are included.  Spaenhauer et al.'s sampling seems best-reproduced by
making the following assumptions about the generic selection function: $V_{\rm
max}\eq17.5$, $V_{\rm min}\eq16$, $A_V\eq1.8$, $\sigma_A\eq0.2$, $M_V\eq1.3$ and
$\sigma_M\eq1.0$. It should be noted that neither $V_{\rm max}$ nor $V_{\rm
min}$ are hard limits but rather soft boundaries derived from the final sample,
while the average extinction is taken from Stanek (1996). Spaenhauer et al.\
used their measured stars to define the reference frame and so their means in
the proper motions are intrinsic. Only the dispersions in the proper motions --
namely $\sigl\eq3.2\nc{\pm}0.1 \masyr$ and $\sigb \eq 2.8\nc{\pm} 0.1 \masyr$ --
carry dynamical meaning.

Finally, Minniti et al.\ (1992) studied stars in two off-axis fields, namely
$(\ell,b)\eq\wMia$ and \wMib. Stars with $R\le16.5$ were pre-selected. Out of
these only the reddest were taken. The limits in colour were chosen to
correspond to the locus of K stars. They estimate that in the two fields ${\sim}
\,10$ and ${\sim}\,30$ per cent are disk stars. The peculiar motion of the Sun
is assumed to be $15.4\kms$ towards $(\ell,b)\eq\window{51}{23}$. The results
are $\vlos\eq45\nc{\pm}10\kms, \siglos\eq85\nc{\pm}7\kms$ for the first field,
and $\vlos\eq77\nc{\pm}9\kms, \siglos\eq68\nc{\pm}6\kms$ for the second field.  
When comparing the predictions of our model with the Minniti et al.\ data, the
following generic selection function is used: $R_{\rm max}\eq16.5$, $R_{\rm
min}\eq12$, $A_R\eq1.5$ and $\sigma_A\eq0.5$. The mean intrinsic magnitude of
the K stars is taken as $M_R\eq-1.5$. The mean extinction $A_R$ is merely a
crude estimate and is open to debate. The parameters used in the generic
selection functions are listed in \tab{select} for all the three windows.

As an illustration of the generic selection functions, we plot an example in
\fig{defex}. These selection functions are of one magnitude width. In other
words, $m_{\rm max}\eq m_{\rm min}\mi1$. The top left panel shows the selection
functions plotted against heliocentric distance. The top right, bottom left and
bottom right panels are for the three windows \wBaa, \wMia, and \wMib,
respectively. In each case, the upper figure shows the number of stars $n(s)$
picked up at a heliocentric distance $s$, namely
 \beq 
	n(s) \propto \eps(s)\, \rho(s)\, s^2,  
\eeq
 and the lower figure shows the density along the line of sight. The important
point is that the selection function has a crucial influence on the observed
kinematics, as it controls where the stars are picked up. As $m_{\rm min}$ is
varied from $13$ to $16$, the observables are dominated by foreground stars,
then bulge stars proper, and finally stars lying behind the bulge. In principle,
the selection function enables us to probe kinematic structure along lines of
sight. Instead of just a single value of a variable such as the line-of-sight
velocity dispersion, there is really a function depending on the magnitude
cut-off.

\section{THE ORBIT LIBRARY}
This section is concerned with the numerical implementation of our extension of
the Schwarzschild algorithm. The first sub-section discusses the choice of the
sampling distribution function, which controls the selection of the regular
orbits in the orbit library. The second sub-section discusses the choice of the
isotropic component.

In this and subsequent sections, the energy $E$ or the Hamiltonian $H(\bw)\eq
\EJ$ are given in units of $G\msun\kpc^{-1}\approx(978\kms)^2$, while the unit 
of the angular momentum is $\sqrt{G\msun\kpc}\approx978\kpc\kms$ with $G$ 
denoting Newton's constant of gravity.

\subsection{The sampling distribution function}
 Since we use the isotropic DF to populate the irregular parts of phase space,
our orbit library contains only regular orbits.  Regular orbits in a rotating
barred potential fill a relatively small volume of phase space around the closed
prograde orbits and a rather larger volume around the closed retrograde orbits.
We ensure that the sampling distribution $\fsa$ is large in these regions by
using a sum of products of a density distribution $\rho(\br)$ of an ellipsoidal
Hernquist model, together with functions $h(\bp)$ of momentum only:
 \beq \label{defsF}
	\fsa=\sum_{i=1}^3A_i\,\rho(\br)\;h_i(\bp).
\eeq
 Here, the non-negative numbers $A_i$ may be chosen for convenience subject to
the condition $1\eq\int{\rm d}^6\bw\,\fsa$.

In the first component of the sampling distribution $\fsa$, $h_1(\bp)$ is
strongly peaked around the momentum of closed prograde orbits. For the second
component, $h_2(\bp)$ is peaked around the velocity of closed retrograde orbits.
For the third component, $h_3(\bp)$ has a broad peak around $\bp\eq0$. We refer
to these components as the prograde, retrograde and hot components,
respectively.

Mathematically, for the hot component, $h$ is given by
 \beq
	h_3(\bp)={1\over(2\upi)^{3/2}\sigma_x\,\sigma_y\,\sigma_z}
	\exp\!\left[{-}\,{p_x^2\over2\sigma_x^2}\mi{p_y^2\over2\sigma_y^2}
	            \mi{p_z^2\over2\sigma_z^2}\right],
\eeq
 where the $\sigma_i$ are specified by \tab{sig_sampling}.

In our coordinate system, the Galactic bar and disk have negative angular
momentum (see \fig{geom}). We define $p_\phi$ to be $\bephi\nc{\cdot}\bp$,
(which is not the momentum conjugate to azimuth $\phi$ but the Cartesian
momentum $\bp$ resolved in the direction of increasing $\phi$) and assume that
on prograde orbits $\bp\nc{\simeq}-v_c\, \bephi$, while on closed retrograde
orbits $\bp\eq v_c\,\bephi$, where $v_c$ is defined by
 \beq
	v_c(R) = 0.25\bigg[1+\Big({0.1\kpc\over R}\Big)^{0.2}\bigg]^{-1}.
\eeq
 For the prograde and retrograde components, $h$ is given by
 \beq\label{defsh12}
	h_{1,2} = {1\over(2\upi)^{3/2}\sigma_R\sigma_\phi\sigma_z}
	\exp\!\left[{-}{p_R^2\over2\sigma_R^2}
		\mi{(p_\phi{\mp}v_c)^2\over2\sigma_\phi^2\!\!}
		\mi{p_z^2\over2\sigma_z^2}\right]\!,
\eeq
 where the plus sign is taken for the prograde component, and the minus sign for
the retrograde component. \tab{sig_sampling} gives the values of the parameters
that appear in \eqn{defsh12}.

The spatial parts of the sampling density~(\ref{defsF}) are defined by
 \beq
	\rho(\br) = {10\over 27\upi }\,
	{1.5\kpc\over m}\,\Big(1+{m\over 1.5\kpc}\Big)^{-3}
\eeq with \beq
	m^2 = x^2+y^2+ (z/0.4)^2.
\eeq
\begin{table}
\caption[]{The velocity dispersions in $\!\kms$ of the three components of the sampling
	distribution function $\fsa$~(\ref{defsF}). \label{tab:sig_sampling}}
\begin{center}
\begin{tabular}{l@{$\;\;$\vline$\;\;$} rrrrr}
\multicolumn{1}{l}{Component} & $\sigma_R$ & $\sigma_\phi$
		& $\sigma_x$ & $\sigma_y$ & $\sigma_z$ \\ \hline 
prograde 	& 60 & 60 &  -  &  -  & 50 \\ 
retrograde 	& 60 & 60 &  -  &  -  & 50 \\ 
hot	 	& -  & -  & 100 & 100 & 60 \\
\hline 
\end{tabular} \end{center}
The dispersions, given in $\kms$, are used in the sampling function and provide
the basis for choosing the orbits in the prograde, retrograde and hot components
of the orbit library. 
\end{table}
 \begin{figure}
\caption[]{Histograms of the distribution of normalized Liapunov exponents for 
	orbits within three narrow ranges of values of the Jacobi energy $\EJ$. 
	The total area under each of the curves is normalized to unity. The 
	dotted vertical lines show our division into regular and irregular 
	orbits at these Jacobi energies. Orbits to the left of the dotted line 
	are regular, orbits to the right are irregular. \label{fig:lambdas}}
\end{figure}

Orbits are followed for ${\sim}\,200$ dynamical times. Over this time, the
Jacobi energy is conserved to typically one part in $10^6$. Liapunov exponents
$\lambda$ are used to distinguish regular from irregular orbits on the principle
that $\lambda\eq0$ for a regular orbit. The process of estimating $\lambda$ is
discussed by Udry \& Pfenniger (1988). In practice, we extrapolate to infinite
time by fitting the estimate $\bar\lambda(t)$ obtained by following the orbit
for time $t$ to $\bar\lambda\eq\lambda\pl b/t$. These unnormalized Liapunov
exponents are converted to normalized ones by multiplying by the orbital time.
This is defined to be the mean time between successive passages through the
plane ${\dot y}\eq0$. \fig{lambdas} shows three histograms of values of
normalized Liapunov exponents $\lambdan$ for orbits in three narrow ranges of
$\EJ$.  In all three cases, there is a sharp peak around $\lambdan\eq0$, which
corresponds to the regular orbits.  After integrating for an infinite time, one
might expect all irregular orbits at any one Jacobi energy $\EJ$ to be
equivalent. Since we only calculate for finite time, the peak corresponding to
the irregular orbits is broadened. The dotted vertical lines in \fig{lambdas}
separate the regular and the irregular orbits. In addition to this, we also
include all orbits whose Liapunov time is greater than five bar rotation times.
This seems reasonable as such orbits do not evolve on the time-scale
(${\sim}\,100$ bar rotation periods)  of interest to us.

The prograde sampling function is used to select $200\,000$ initial conditions.
These give rise to only $5\,713$ regular orbits. Similarly, $100\,000$ initial
conditions selected from the sampling distribution of the hot component give
rise to $5\,512$ regular orbits.  Only $50\,000$ initial conditions selected
from the sampling function of the retrograde component give rise to $10\, 943$
regular orbits.  The final orbit library contains $\Nr\eq 22\, 168$ regular
orbits. The library records the probabilities of each orbit being in any given
cell and the time-averaged sampling density.

\fig{u_mass} shows the density of the library's orbits in a convenient
projection of orbit space.  In a conventional Lindblad diagram for an
axisymmetric galaxy, the angular momentum $L_z$ of orbits is plotted
horizontally, and their energy is plotted vertically. The top panel of
\fig{u_mass} is a modified Lindblad plot for our system, in which {\it
orbit-averaged\/} values of $L_z$ and energy are plotted horizontally and
vertically. These averages are effective rather than classical integrals. Since
the Hamiltonian satisfies $H\eq E-\Omega L_z$, the classical integral $H$ is a
linear combination, $\langle E\rangle\mi\Omega\langle L_z\rangle$ of the
effective integrals. We choose to plot $\langle E\rangle$ rather than $H$
because the former is a truer guide to an orbit's physical size than the latter.
In a classical Lindblad diagram, the allowed region $|L_z|\nc{\le}L_{\rm c}(E)$,
where $L_{\rm c}(E)$ is the angular momentum of a circular orbit of energy $E$,
tapers as one descends to smaller values of $E$. In \fig{u_mass} this tapering
has been largely suppressed by plotting horizontally not $\langle L_z\rangle$
but $\langle L_z\rangle/L_{\rm c}$, where $L_{\rm c}$ is not the angular
momentum of a circular orbit, which does not exist in a barred galaxy, but is
defined to be (in the units given above)
 \beq\label{defLc}
	\Lc\big(\langle E \rangle\big) = 
	\frac{1}{600} \big(\langle E \rangle + 0.25\big)^4.
\eeq

\begin{figure*}
\makebox[\hsize]{	\epsfxsize=0.75\hsize
}
\makebox[\hsize]{	\epsfxsize=0.69\hsize
}
\caption[]{{\bf Top panel:} The density of orbits in the library of regular 
	orbits is colour-coded, with black the lowest density. 
	Scaled, time-averaged angular momentum is plotted horizontally 
	and time-averaged energy vertically. The red curve is the contour 
	$H\eq\Peff^{\rm rim}$, above which orbits are not confined by the 
	Hamiltonian and the isotropic part of the DF, $\fe$, is set to zero.
	Individual orbits are marked by white dots. 
	{\bf Lower panels:} representative orbits. The location of each orbit 
	in the top panel is indicated by the letter to the right or the orbit 
	and the $(\avet{E},\,\avet{L_z})$ coordinates on top of it. Orbits are 
	followed for 200 dynamical times. \label{fig:u_mass} }
\end{figure*}

Since the Galactic bar has a negative pattern speed, prograde orbits lie on the
left hand side of \fig{u_mass} and contours of constant $H$ slope from top left
to bottom right. The red curve in \fig{u_mass} is the contour for $H\eq
\Peff^{\rm rim}$. The region of strong chaos associated with corotation is
evident in the white sea that cuts the $\langle E\rangle$ axis between $-0.035$ and
$-0.075$. Within this sea there is a long thin island of regularity. The panels
marked C1 and C2 show two orbits within this island. One is trapped around the
standard maximum of $\Peff$ at $L_4$. The other is trapped around the
non-standard maximum of $\Peff$ at $(\pm 2.9,\pm 2.2)\kpc$. At smaller values of
$\langle E\rangle$, a ridge of orbits trapped around the prograde, bar-supporting
$x_1$ family is apparent near the left-hand edge of \fig{u_mass}. The panels
marked A, B1 and B2 in the lower half of \fig{u_mass} show representative orbits
from this region. Panel D shows a prograde orbit at the largest value of
$\langle E\rangle$ plotted in \fig{u_mass}. This is a typical disk orbit.

On the right-hand, retrograde side of orbit space, orbits of the $x_4$ family
occupy a region of regularity that extends, unbroken, from the smallest to the
largest energies. The panels labelled K, L and M show representative orbits from
this region.

In \fig{u_mass}, colour shows the density of orbits, while white dots show the
orbits themselves. Many sharp chains of dots can be discerned: these mark the
paths of stable resonances.

\subsection{The isotropic component} \label{sec:fiso}
 The isotropic DF can be conveniently represented as a linear combination of
some basis functions -- the coefficients in this expansion are the free
parameters that then determine $\fe$.  We have employed second-order
basis-splines (see e.g., Stoer \& Bulirsch 1980). Hence,
 \beq\label{fctowc}
	\fe(\bw)=\sum\limits_{i=1}^{\Ni}\,\wi_i\,
	B_i\big(H(\bw)\big),
\eeq where $\Ni\eq1000$ and
\beq\label{equ_BH_def}
	B_i(\EJ) = {k_i\over\Delta \EJ} \times \bec
	  \EJ\mi\EJ_{i{-}1}& for $\EJ\nc\in[\EJ_{i{-}1},\,\EJ_i]$, \\[.5ex]
	  \EJ_{i{+}1}\mi\EJ& for $\EJ\nc\in[\EJ_i,\,\EJ_{i{+}1}]$, \\[.5ex]
	  0 & otherwise. 
\eec \eeq
 Here, $\Delta\EJ$ denotes the grid spacing, taken to be constant, while the
$k_i$ are constants that enforce some chosen normalization. The basis functions
$B_i\big(H(\bw)\big)$ can be interpreted as building blocks containing all
orbits with Jacobi integral $\EJ\eq H(\bw)$ for which $B_i(\EJ)\nc{>}0$.

As mentioned in Section~3.1, outside corotation, irregular orbits can escape
to infinity. Since these irregular orbits are occupied via $\fe$, we must
restrict $\fe$ to values of $\EJ$ of orbits that cannot cross the potential rim
at corotation. This means that (i) $\fe$ is non-zero only for $\EJ\nc<
\Peff^{\rm rim}$, and (ii) at fixed $\EJ\nc<\Peff^{\rm rim}$ refers only to
orbits inside corotation. Thus, strictly speaking equation (\ref{fctowc}) 
for $\fe$ contains
two Heaviside functions, one ensuring $\EJ\nc<\Peff^{\rm rim}$ and the other
$|x|\nc<x_{\rm rim}$, where $\Peff^{\rm rim}\eq\Peff(x_{\rm rim},0,0)$.  
For simplicity, we have suppressed these Heaviside functions.

Computing moments of the isotropic DF, $\fe$, reduces to computing them for the
$B_i$. For the mass density, for example, \beq\label{rho-iso}
	\rhoi(\br)\eq \Mb \sum_i\wi_i\pe_i(\br),
\eeq where 
\beq\label{equ_rhoc}
	\pe_i(\br)\equiv\!\int\!\d^3\bv\,B_i(\EJ) =
		\frac{4\sqrt{2}\,k_i}{15\,\Delta\EJ} 
		\left[D_{i-1}^5 {-} 2 D_i^5 {-} D_{i+1}^5\right]
\eeq
with $D_i^2\id\EJ_i\mi \Peff(\br)$. The $k_i$ are determined by normalizing 
$\pe_i$ such that
 \beq
	1 = \int \d^3\br\;\pe_i(\br).
\eeq
This integral over real space is done numerically by the Monte-Carlo method. 

\begin{table}
\caption[]{Labelling of the velocity moments in the merit function.
	\label{tab:label}}
\begin{center} \begin{tabular}{ll}
$i$		&	\qquad$\omega_{ji}$			\\ \hline
$1\dots3$	& $\ave\vlos,\,\ave\mul,\,\ave\mub$		\\
$4\dots6$	& $\ave{\vlos^2},\,\ave{\mul^2},\,\ave{\mub^2}$	\\
$7\dots9$	& $\ave{\vlos\mul},\,\ave{\vlos\mub},\,\ave{\mul\mub}$ \\ \hline
\end{tabular} \end{center}
Not all these components of the velocity dispersion tensor are
available for all the windows. The most complete information is known for
Baade's Window -- namely, \vlos, \siglos, \sigl\ and \sigb. \end{table}

\begin{figure*}
	\epsfxsize=\hsize\epsfbox[62 570 461 767]{\figdir/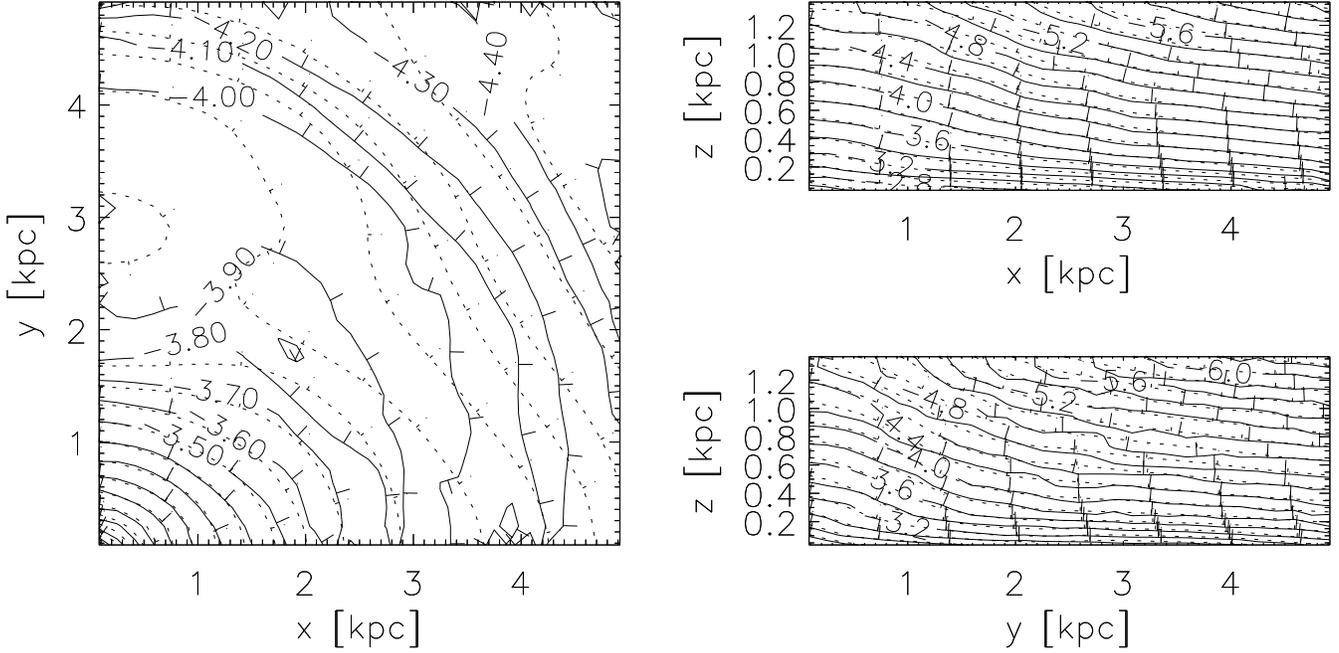}
\caption[]{Logarithmic contours of the projected mass in the principal planes 
	of the Binney et al.\ (1997) model (dotted lines) are compared with
	those of our 
	dynamical model (full lines). The units are such that the total mass 
	in the $10\kpc\times10\kpc\times2.8\kpc$ box is unity.  \label{fig:proj}}
\end{figure*}

\section{THE MERIT AND PENALTY FUNCTIONS}
The combined mass density of the orbits must match the original density $\rhoo$
used to create the potential (see Section 3.1). To this end we minimize the
merit function
 \beq
	\Qden	=\sum_j \left[\frac{\Delta_j}{\rho_{{\rm norm}\,j}} \right]^2,
\eeq 
 where the sum extends over all cells and
\beq\label{defstriang}
	\Delta_j = {\Mo_j\over\Mb}-\left[\sum\limits_{i=1}^{\Ni}\wi_i\,\pe_{ij}
	+ {1\over\Ns}\sum\limits_{i=1}^{\Nr}\wr_i\,\pr_{ij}\right],
\eeq
where $\Mb\eq5.18\nc\times10^{10}\msun$, while $\Mo_j$ is the mass contained in
the $j$th cell of our mass model according to BGS. $\pe_{ij}$ is the integral of
$\pe_i(\br)$ [equation~(\ref{equ_rhoc})] over cell $j$. Because the $B_i$ are normalized 
to unit phase-space integral, $\pe_{ij}$ is just the probability for a star 
whose phase-space coordinates are drawn from $B_i\big(H(\bw)\big)$ to be found 
in cell $j$ at any time. Equivalently, $\pr_{ij}$, defined by \eqn{pij}, is the 
probability that a star on the $i$th regular orbit is found in cell $j$ at any 
time. The weighting of the cells, $\rho_{{\rm norm}\,j}$, is chosen to be
 \beq
	\rho_{{\rm norm}\,j}=\frac{\rhoo_j}{\sqrt{N_{{\rm sam}\,j}}},
\eeq
 where $N_{{\rm sam}\,j}$ is the number of building blocks contributing to grid
cell $j$. The effect of this weighting scheme is a minimization of relative
errors in the density modified by the Poisson error expected from the sampling.

The kinematics of our model are given by the matrix $\omega^{\rm m}_{ji}$ with
$i\eq0,\dots,9$ representing the $i$th velocity moment in window $j$, as defined
in \tab{label}. Appendix B gives details of the calculation of the moments. The
contribution to the merit function of the kinematic constraints is
 \beq
	\Qkin =	\sum_j\sum_{i=1}^{9}\gamma_{ji}\left(
		{\omega^{\rm m}_{ji}}-{\omega^{\rm o}_{ji}}\right)^2,
\eeq
 where $\gamma_{ji}$ is a weighting matrix and the superscripts m and o refer to
the model and observed quantities, respectively. In some windows only a subset
of the moments is available. For all three of the constraint windows (described
in Section 3.2), line-of-sight velocity \vlos\ and dispersion \siglos\ are
known. Only in Baade's Window are the dispersions \sigl\ and \sigb\ also
available.

The problem posed by the minimization of the merit function is very
ill-conditioned, so regularization is necessary. This is achieved by enforcing a
smoothness constraint -- `neighbouring' orbits should have `similar' weights. A
suitable penalty function is the mean-square value of the second derivative of
the logarithm of the weights with respect to some distance -- this vanishes when
the dependence of weights on distance is a power law.  For orbits belonging to
the isotropic DF, a suitable distance is provided by $\EJ$, and the penalty
function reads
 \beq 
	\Piso=\frac{1}{\Ni}\sum_{i=2}^{\Ni-1}\left[
	\frac{\ln\wi_{i-1} - 2\ln\wi_i + \ln\wi_{i+1}}{(\Delta\EJ)^2}
	\right]^2.
\eeq
 For the regular orbits, the distance measure is provided by a set of effective
integrals (c.f.\ Merritt \& Fridman 1996, Zhao 1996). Specifically, we employ
the time averages $I_1\id\avet{p_z^2}$, $I_2\id\avet{L_z}$ and $I_3\id\avet E$ 
-- note that $I_3$ is the mean orbital {\em energy\/}, not the value of $\EJ$.
The penalty function for regular orbits then reads 
\beq
	\Preg=\frac{1}{\Nr}\sum_{i=1}^{\Nr}\left[
	\frac{\ln\wr_i - \ave{\ln\wr}_{A(i)}}{\ave{d}_{A(i)}}\right]^2.
\eeq
 Here, $A(i)$ is a neighbourhood of orbit $i$ and $\avet{}_{A(i)}$ denotes the
average within this neighbourhood. $A(i)$ is defined to consist of the 8 orbits
closest to orbit $i$. The distance, $d_{ij}$, between orbits $i\nc{\neq}j$ is
defined by
 \beq
	d_{ij} \equiv \sum_{k=1}^3\left[
	\frac{I_{ki}-I_{kj}}{\sigma_{I_k}}\right]^2,
\eeq
 where $\sigma_{I_k}$ is the dispersion of $I_k$ over all the regular orbits.

The final quantity to be minimized, $\ov{Q}$, is a linear combination of the
density and kinematic merit functions, $\Qden$ and $\Qkin$, and the penalty
functions $\Preg$ and $\Piso$. The final numerical factors in this linear 
combination were chosen as follows. The relative weight for $\Qkin$ was chosen
to be as small as possible without increasing the deviation from the observed 
kinematics by more than the observational uncertainty ($1\sigma$). Similarly, 
the relative weight for the penalty functions were chosen as large as possible
without worsening the fit to the mass density by more than  2 per cent overall.

To enforce the non-negativity of $f$, we substitute the weights $\wi_j$ and
$\wr_j$ by $\beta^{\rm c}_j$ and $\beta^{\rm r}_j$ in the following way:
 \bsq \label{beta} \ben
	\beta^{\rm iso}_j	&=&	\ln\wi_j	\\
	\beta^{\rm reg}_j	&=&	\ln\bigg[\wr+\ov\fsa(\bJ_j)^{-1}
				\sum\limits_{i=1}^{\Ni}\,\wi_i\, 
				B_i\big(H(\bJ_j)\big)\bigg].
\een\esq 
 Hence, $\beta^{\rm reg}\to-\infty$ as $\wr$ tends to the lowest value
compatible with the positivity constraint~(\ref{truecon}). Minimizing $\ov{Q}$
with respect to $\{\beta^{\rm iso}_j,\,\beta^{\rm reg}_j\}$ always results in
a physical model. Unfortunately, two attractive properties of the original
optimization problem are lost in the transition from the $w$'s to the $\beta$'s:
$\Qden$ no longer depends linearly on the variables, and the boundedness of the
solution is not guaranteed (as the $\beta$'s can diverge).

Due to the great number of orbit weights to be determined, the memory
requirement of the adopted optimization algorithm must not increase with the
number of unknowns faster than linearly. This excludes fast schemes such as
non-negative least square fitting (e.g., Zhao 1996).  Another popular choice,
the iterative Richardson-Lucy method (Newton \& 
Binney, 1984; Statler, 1987), is not applicable to our problem because its
kernel is not positive definite.
 Our final choice fell on the conjugate gradient algorithm (e.g., Stoer \&
Bulirsch 1980; Press et al.\ 1988). It satisfies the stringent memory
requirements and is an improvement on the steepest descent method in so far as
the directions in which it does its line minimizations are conjugate to each
other.

\begin{figure}
	\epsfxsize=\hsize\epsfysize=.91\hsize
	\epsfbox[289 359 489 567]{\figdir/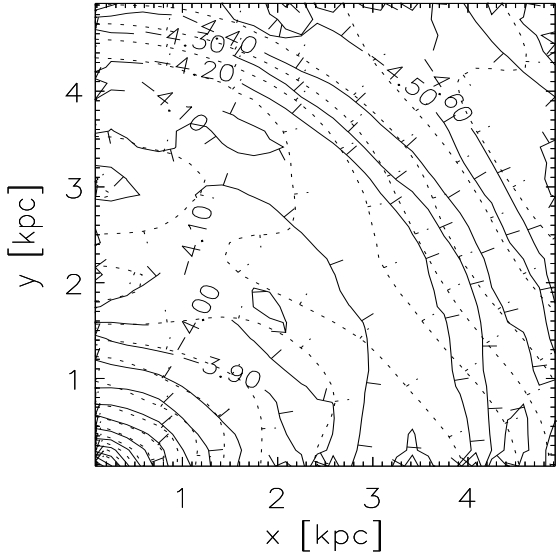}
\caption[]{The mass within 0.14 kpc of the Galactic plane is shown for the 
	Binney et al.\ (1997) input density (dotted contours) and for  
	the present model (full contours). While the model can reproduce the 
	mass in the inner parts very well it fails to reproduce the 
	over-densities on the $y$-axis that are most likely due to symmetrized 
	spiral arms. \label{fig:unproj}}
\end{figure}
\begin{figure*}
	\epsfxsize=\hsize\epsfbox[25 569 525 764]{\figdir/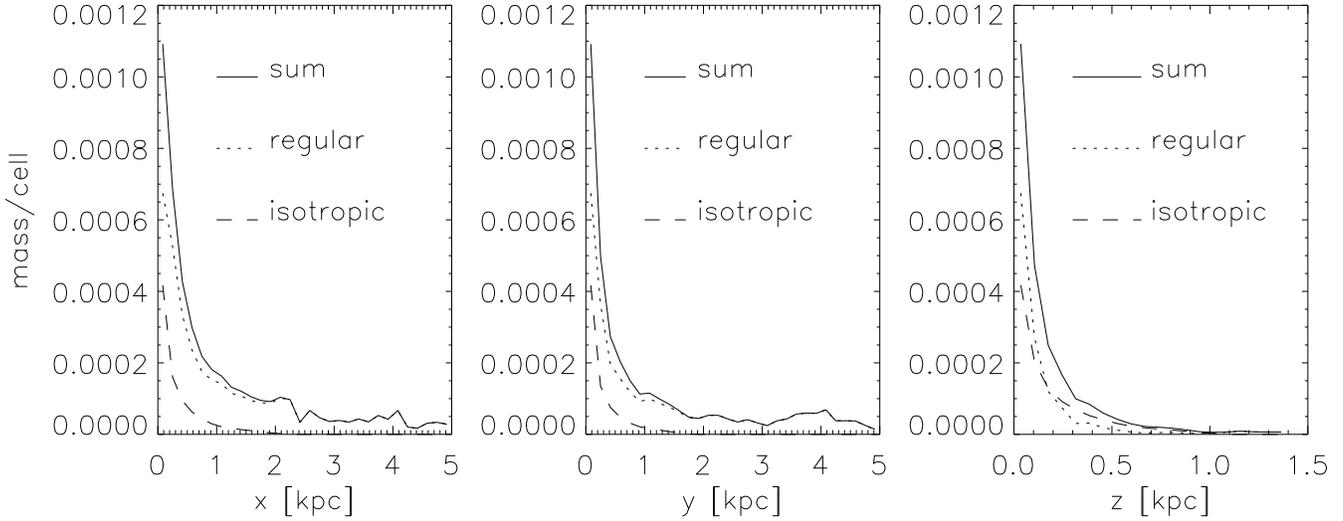}
\caption[]{The mass profiles of  our model are shown along 
	the $x$, $y$ and the $z$-axes. The isotropic DF contributes two-thirds  as 
	much mass as the regular component near  the centre. Along the 
	$x$-axis the contribution from the isotropic DF decreases rapidly, 
	while along the $z$-axis the isotropic component becomes steadily more
	important until it dies out near $z \sim 1.2$. 
	Relatively few regular orbits pass through the minor axis, so
	the discrepancies between our model and the input density are
	substantial further than $1.2\kpc$ down the $z$-axis.
	\label{fig:massprof} }
\end{figure*}

\section{RESULTS}
This section describes our dynamical model of the Milky Way's bar. Section 6.1
describes how the model reproduces the density and kinematic constraints.
Section 6.2 provides kinematic predictions of the model, while Section 6.3
discusses the phase-space structure.

\subsection{The constraints}
\subsubsection{The density}
 The full contours in \fig{proj} show the density of our final dynamical model
projected onto the three principal planes. The broken contours show the
corresponding projections of the input density of Binney et al.\ (1997).
Overall, the fit between the two density distributions is good, with the average
discrepancy $\sim\, 10 \%$. The biggest contribution to this error comes from
the innermost two layers of cells close to the Galactic plane, and particularly
around corotation on the $y$-axis. Outside these areas, the error is very much
smaller.  The model reproduces both the axis ratio and the detailed shapes of
the contours in the $(x,z$) and $(y,z$) planes very well out to the limits of
the box, although there are some small discrepancies on the $z$-axis -- see
below.

\fig{unproj} highlights the discrepancies between the model and input densities
in the ($x,y$) plane by contouring for each model the mass within $0.14\kpc$ of the
plane.  The two sets of contours agree well within corotation, but significant
discrepancies occur further out. These discrepancies are not worrying for two
reasons.  First, obscuration is at its worst in the Galactic plane, and in the
region one cannot confidently deproject the surface photometry.  Second, the
deprojection algorithm of Binney et al.\ assumes that all structures are
eight-fold symmetric with respect to the principal planes of the bar. Spiral
arms are less symmetric and spurious features will arise when one attempts to
deproject a spiral distribution under the assumption of eight-fold symmetry.  
Indeed, Binney et al.\ show that features remarkably like the `observed' the
density maxima along the $y$-axis arise when the Binney et al.\ algorithm is
used to deprojection a four-armed spiral distribution. Englmaier \& Gerhard
(1998) provide further evidence for this interpretation by showing that better
fits to the observed longitude--velocity diagrams for HI and CO are obtained
when the flow of gas in the inner Galaxy is calculated in a potential that
includes contributions from the density maxima. Moreover, the
contributions from the maxima induce a four-armed rather than a predominantly
two-armed spiral in the gas. Hence several lines of evidence indicate
that the density maxima in the deprojected disk are artifacts
associated with four spiral arms. Such spiral arms inevitably lie beyond the
reach of our modelling technique.

\fig{massprof} shows density profiles of the model along the $x$- and $z$-axes.
At the centre the isotropic component contributes over 30 per cent of the
mass.  Within the plane this component diminishes rapidly in importance as
one moves away from the centre.  Along the $z$-axis, by contrast, the
isotropic component gains in importance as one moves from the centre because
many of the orbits passing through the $z$-axis are irregular.  Beyond $\sim
1.2\kpc$, the isotropic component is unable to contribute to the density
because orbits inevitably have Jacobi energies in excess of $\Phi_{\rm
eff}^{\rm rim}$, and the discrepancies between our model and the deprojected
photometry increase.

\begin{table}  
\caption[]{Kinematic data used to constrain the model.
	\label{tab:kin-constraint}}
\begin{center}
\begin{tabular}{llr@{}llr}
	($\ell,b$)&Quantity & \multicolumn{2}{c}{observed} & Model & Ref
\\ \hline
	\wBaa	& \vlos		&   4&$\pm8$	& 4	& Sh	\\
		& \siglos	& 113&$\pm5$	& 115	& 	\\[1ex]
	\wBaa	& \siglos	& 120&		& 114	& Sp	\\
		& \sigl		& 3.2&$\pm0.1$	& 3.6	& 	\\
		& \sigb		& 2.8&$\pm0.1$	& 2.8	& 	\\[1ex]
	\wMia	& \vlos		&  45&$\pm10$	& 45	& M	\\
		& \siglos	&  85&$\pm7$	& 80	& 	\\[1ex]
	\wMib	& \vlos		&  77&$\pm7 $	& 75	& M	\\
		& \siglos	&  68&$\pm6$	& 95	&
\\ \hline
\end{tabular} \end{center}
 The units are $\kms$ for velocity dispersions and $\masyr$ for proper-motion
dispersion ($\sigma_\ell$ and $\sigma_b$). The last column is the mnemonic for
the observers (see \tab{select}). The biggest discrepancies occur at \wMib,
where our dynamical model has too much dispersion and not enough streaming.  
\end{table}

\begin{table*}	
\caption[]{Model predictions for unmeasured quantities.\label{tab:kin-predict}}
\begin{tabular}{lr@{}lr@{}lr@{}lrr@{}lr@{}lr@{}lr@{}lr@{}ll}
$(\ell,\,b)$ & 
	\multicolumn{2}{c}{\vlos} &
	\multicolumn{2}{c}{\mul} &
	\multicolumn{2}{c}{\mub} &
	\multicolumn{1}{c}{\siglos} &
	\multicolumn{2}{c}{\sigl} &
	\multicolumn{2}{c}{\sigb} &
	\multicolumn{2}{c}{$C_{s\ell}$} &
	\multicolumn{2}{c}{$C_{s b}$} &
	\multicolumn{2}{c}{$C_{\ell b}$} & Ref	\\ \hline
\wBaa	&    &  &  0&.26&--0&.01&   &3&.5&2&.8&--0&.05&  0&.03&  0&.04&Sh\\
\wBaa	&   3&.8&  0&.25&--0&.01&   & &  & &  &--0&.05&  0&.03&  0&.04&Sp\\
\wMia 	&    &  &--0&.85&--0&.22&   &6&.0&3&.4&  0&.52&  0&.05&--0&.14&M \\
\wMib 	&    &  &  0&.70&--0&.10&   &4&.4&2&.1&  0&.08&--0&.05&  0&.01&M \\[1ex]
\wTT	&  77&  &  0&.87&  0&.29& 78&3&.1&3&.0&  0&.31&--0&.01&  0&.15&TT\\
\wBla	&   8&  &--0&.16&--0&.09&134&3&.5&3&.0&--0&.11&  0&.02&  0&.04&Bl\\
\wBlb	& --8&  &--0&.20&  0&.01&112&4&.4&2&.6&--0&.10&--0&.01&  0&.03&Bl
\\ \hline
\end{tabular}

\smallskip
\begin{minipage}{130mm}
 Units are $\masyr$ for proper-motions and $\kms$ for velocities. The $C_{ij}$
are the dimensionless correlation coefficients of the observable velocity dispersion tensor,
that are defined by \eqn{correl}. The first four windows are those used to constrain the
model (data used for that purpose are given in \tab{kin-constraint} and are
omitted here), the fourth window has been observed by Tiede \& Tendrup~(1997),
and the last two by Blum et al.\ (1997). The line of sight streaming for the
Blum et al.\ fields has not been corrected for the reflex motion of the Sun to
enable comparison with his data.
 \end{minipage}
\end{table*}

\subsubsection{The kinematics}
 It is essential to apply both photometric and kinematic constraints to obtain
plausible models of the inner galaxy. The dynamical model is required to
reproduce the data provided by Sharples et al.\ (1990) and Spaenhauer et al.\
(1992) in Baade's Window, together with the observations of Minniti et al.\
(1992) at \wMia\ and \wMib.  Note that we maintain a distinction between the
Sharples et al.\ (1990) and the Spaenhauer et al.\ (1992) data-sets, as the
model is viewed at the same window through different selection functions.
\tab{kin-constraint} shows how the model fares. In Baade's Window, the
line-of-sight streaming velocity, \vlos, and velocity dispersion, \siglos,
together with one of the proper-motion dispersions \sigb\, are well reproduced. 
The remaining proper-motion dispersion \sigl\ is higher than measured by
Spaenhauer et al.\ (1992). This is because our orbit library, and hence our
model, probably contains too many retrograde orbits.  Turning to the Minniti et
al.\ (1992) fields, the streaming velocity and the dispersions at \wMia\ are
reproduced to within the error bars. The more distant window at \wMib\ is more
difficult to get right. One worry is that disk contamination is likely to be
severe in this outer window, which may mean that Minniti et al.'s (1992) results
need correction. This explanation is consistent with the fact that our model has
a higher dispersion and a lower streaming velocity than are suggested by the
data. Overall, though, \tab{kin-constraint} encourages us in the belief that 
our dynamical model is a good representation of the inner Milky Way and that it 
is useful to make some kinematic predictions from the model.

\subsection{Kinematic predictions}
 Let us stay for the moment with our constraint fields. In Baade's Window, the
means in the proper motions $\mul$ and $\mub$ are not given by Spaenhauer et
al.\ (1992), but it may become possible to recover them at some time in the
future. The
mixed components of the velocity dispersion tensor are also thus far unmeasured,
although there is preliminary claim of a measurement of the vertex deviation
from a small sample by Zhao, Spergel \& Rich (1994). In the Minniti et al.\
(1992) fields, only line-of-sight quantities are available. \tab{kin-predict} 
presents the predictions of our model for all the unmeasured quantities. [We 
have not given results for \wMib\ because our dynamical model does not reproduce
the existing data there.] We also compute the correlation coefficients between 
the observable velocity dispersions
 \beq \label{correl}
	C_{s \ell} \equiv {\sigrl \over \siglos\,\sigl},	\qquad
	C_{s    b} \equiv {\sigrb \over \siglos\,\sigb},	\qquad
	C_{\ell b} \equiv {\siglb \over \sigl  \,\sigb}.
\eeq
 Zhao, Spergel \& Rich (1994) suggest that $C_{s\ell}$ is a useful diagnostic of
bulge triaxiality. At Baade's Window, this quantity vanishes for a steady-state
axisymmetric density distribution. The predictions for these correlations are
presented in the tables. They are, of course, measures of the misalignment of
the principal axes of the velocity dispersion tensor with the ($s,\ell,b$) axis
set. In Baade's Window, the  dispersion tensor has principal semi-axes in the
ratio $0.83:1:0.77$, with the longest axis pointing almost 
in the $\bel$ direction. Hence, Baade's Window is a rather poor place 
to look for the signature of triaxiality. Our barred model has a dispersion 
tensor whose alignment is almost the same as that of an oblate axisymmetric 
model! In the Minniti et al.\ \wMia\ field, the dispersion tensor is very 
strongly anisotropic, with semi-axes in the ratio $0.25:1:0.50$. The long axis 
points almost in the $\bel$ direction and has the high value of $238\kms$. In 
the model, there are more and more retrograde stars picked up as one moves 
further from the Galactic Centre. The velocity dispersion in the longitudinal 
direction rises as the prograde and retrograde stars become present in 
almost equal numbers.

Now, let us see how our model fares in comparison with data for two new fields. 
Tiede \& Terndrup (1997) present the results of a study of 189 stars in a field
at $(\ell,b)\eq(8.4^\circ,\,-6.0^\circ)$. The selection criteria are described
in detail in their paper but are not simple to reproduce. A crude approximation
to their selection procedure is to use the generic selection function
(\ref{generic}) with parameters $m_{\rm max}\eq17.0$ and $m_{\rm min}\eq12$.  
Tiede \& Terndrup provide values for the mean extinction $A\eq1.1$ and its
dispersion $\sigma_A\eq0.2$. They measure the line-of-sight dispersion $\siglos$
of their sample to be $75\nc\pm1\kms$. \tab{kin-predict} shows the predictions
of our model. Good reason for believing in the model's reliability within
corotation is that this value of the dispersion is very reproduced. The
diagonalised tensor has semi-axis ratios $0.42:1:092$ and is rather strongly
misaligned with the ($s,\ell,b$) axis set, as the large correlation $C_{s\ell}$
indicates.

Blum et al.\ (1994) studied stars in two fields very close to the Galactic
Centre at \wBla\ and at \wBlb. The stars comprising the sample were selected in
a manner that tried to exclude disk stars. It is not so easy to reproduce their
selection procedure. A crude approximation is to take the selection function as
unity for all heliocentric distances $s$ satisfying $6\kpc\nc< s \nc< 10\kpc$.
From their data, Blum et al.\ deduce that $\vlos\eq14\nc\pm23 \kms$ and
$\siglos\eq128\pm14\kms$ at \wBla\ and that $\vlos\eq-75\nc\pm24\kms$ and
$\siglos\eq127\pm17$ at \wBlb. The predictions of our model for these two new
fields are reported in \tab{kin-predict}. Again, there is the reassuring
circumstance that both the dispersions are reproduced to within the errors. One
of the streaming velocities is recovered to within the error bars, but one is
not. At \wBla\ the dispersion tensor has semi-axes
$(141,\,126,\,112)\kms$, while at \wBlb\ the semi-axes are $(111,\,
167,\, 100)\kms$. In both cases the principal axes are not strongly
misaligned with the $(s,l,b)$ coordinate directions, so, as in Baade's Window,
triaxiality will be hard to establish unambiguously in these fields.

\begin{figure}
	\epsfxsize=\hsize \epsfbox{\figdir/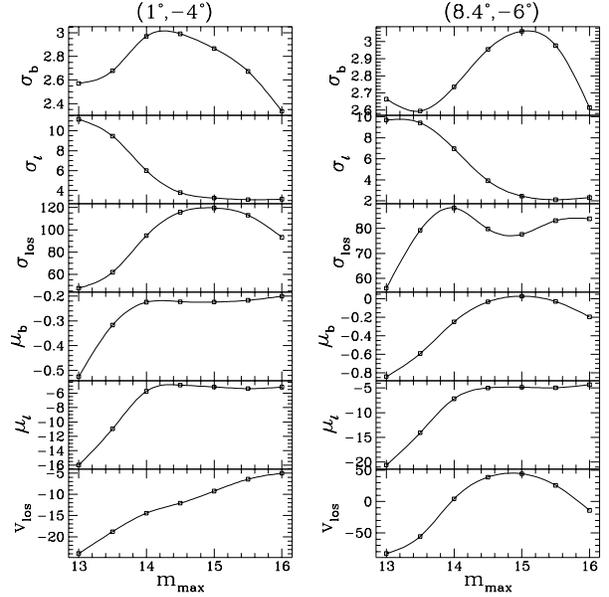}
\caption[]{The left panel gives the kinematic predictions at Baade's Window 
	\wBaa\ and the right panel at the field studied by Tiede \& Terndrup 
	\wTT. Here, the generic selection function is used with $m_{\rm min}\eq
	m_{\rm max}\mi1$. \label{fig:predictBW}}
\end{figure}

\begin{figure}
	\epsfxsize=\hsize \epsfbox{\figdir/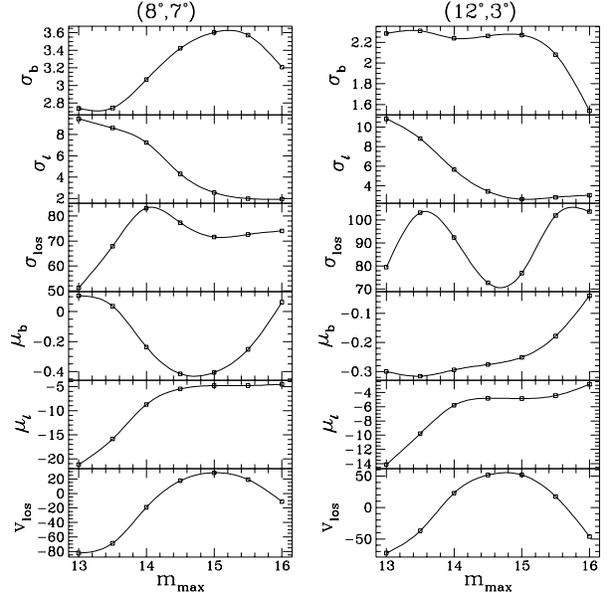}
\caption[]{Kinematic predictions at Minniti et al.'s windows. The left panel 
	shows results for \wMia, the right panel for \wMib. Again, the generic 
	selection function is used with $m_{\rm min}\eq m_{\rm max}\mi1$.
	\label{fig:predictM}}
\end{figure}

Finally, \figsto{predictBW}{predictM} show the kind of kinematic data that
may become available in the very near future. Here, we have imagined that
our dynamical model is observed through four windows using the generic
selection function (\ref{generic}) with a width of one magnitude. We assume
that the extinction and intrinsic magnitude of the stellar population
vanish, so $m_{\rm max}$ is in effect the maximum distance modulus in the
sample.  It is apparent from \figsto{predictBW}{predictM} that a wealth of
additional information is uncovered when the kinematics is studied as
a function of apparent magnitude.

As one application of the figures, let us examine how to improve the slight
discrepancies between the model and the observations in the constraint fields.
The quantity $\sigb$ is too large by a factor of ${\sim}\,5$ per cent in Baade's
Window. From \fig{predictBW}, it is evident that this can be corrected by
slightly increasing the faint cut-off, as the curve of $\sigb$ versus $m_{\rm
max}$ falls for $m_{\rm max}\nc>14.5$. If, for example, either the total
extinction or the faint cut-off is less severe than we have assumed, then
$\sigb$ is lowered to give better agreement with the Spaenhauer et al.\ data.
The quantity $\sigl$ is also slightly too large in our dynamical model, but
\fig{predictBW} makes it clear that increasing the faint magnitude cut-off makes
little difference to its value. Some of the properties of the curves in
\figsand{predictBW}{predictM} are readily explained. The line-of-sight streaming
\vlos\ typically gives a one-humped curve, as it is greatest when stars are
picked up by the selection function at roughly the tangent point. The
line-of-sight dispersion $\siglos$ can sometimes give complex curves, as in the
\wMib\ field of Minniti et al. Here, as we move along the line of sight, the
selection function picks up first a mixture of prograde and retrograde stars in
the near side of the disk, then mainly prograde stars in the bar proper, and
finally prograde and retrograde stars in the far side of the disk. This causes
$\siglos$ to rise, then fall, and then rise again.

Clearly, as sample sizes increase, it will become increasingly inappropriate to
characterize the kinematics of a given field by just mean velocities and
dispersions. Even observations of external galaxies can now deliver other
measures of the line-of-sight velocity distribution (LOSVD), such as the
Gauss-Hermite coefficients $h_3$ and $h_4$. Still richer information should be
available in the case of the Milky Way because, by varying $m_{\rm min}$ and
$m_{\rm max}$, we can probe the LOSVD at different points along the line of
sight. More sophisticated data sets will surely resolve much of the degeneracy
that now plagues dynamical models.

\begin{figure}
\caption[]{The distribution function of the regular component of the final 
	model is shown projected into the $\avet{L_z}/\Lc-\avet{E}$ plane. The 
	phase-space density is largest in the bottom left corner of the figure, 
	which is occupied by small, nearly harmonic box orbits. At higher 
	energies the density peaks along the ridge of the prograde $x_1$-orbit 
	family. \label{fig:f_df}}
\end{figure}

\begin{figure}
\caption[]{The distribution of the regular component's mass within the 
	$\avet{L_z}/\Lc-\avet{E}$ plane. The colour of each cell encodes the sum
	of the weights assigned to the orbits that lie in that cell.
	\label{fig:f_mass}}
\end{figure}

\begin{figure*}
\centerline{ \epsfxsize=120mm \epsfbox[25 175 585 498]{\figdir/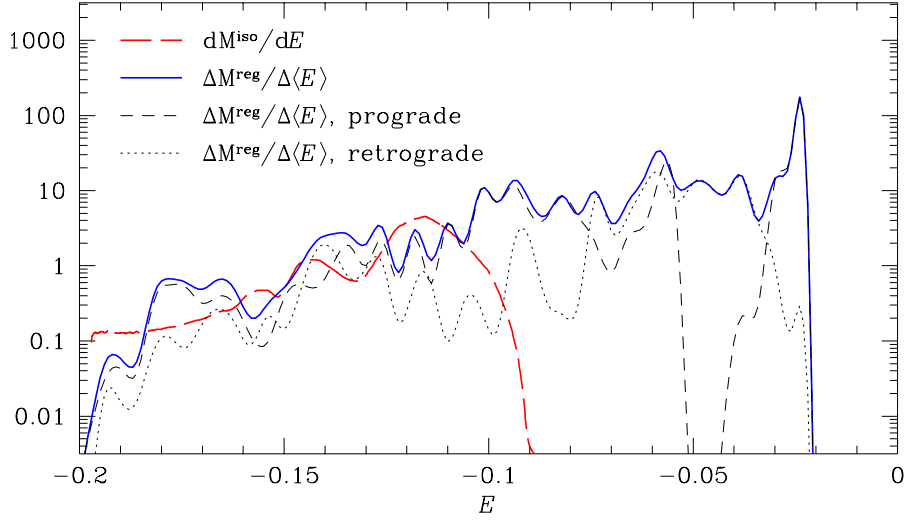} }
\caption[]{The differential mass distribution of the isotropic and the regular
	component. The regular part is further split into prograde and 
	retrograde orbits. The isotropic component stops rather abruptly at 
	$E\eq{-}0.09$ since it is confined to inside corotation. 
	\label{fig:dmde} }
\end{figure*}
 
\subsection{The distribution function}
 \fig{f_df} shows the distribution function of the regular component of the
final model projected onto a plane whose axes are averaged energy $\avet{E}$ and
averaged angular momentum $\avet{L_z}$, normalized by the reference value,
$\Lc$, that is defined by \eqn{defLc}. The phase-space density is highest at
bottom left, and decreases with increasing energy for both prograde and
retrograde orbits. These trends are in the same sense as the variation of $\fe$,
which decreases monotonically as $H$ increases from bottom left to top right.

\fig{f_mass} shows the distribution of the mass of the regular component
within the $(\avet{L_z},\,\avet{E})$ plane: the colour of each pixel encodes
the sum of the weights $\wr$ assigned to orbits whose effective integrals
place them within that pixel. [\Eqn{rhofromrho} shows that the mass on an
orbit is proportional to its weight.] Comparing \figsand{f_df}{f_mass} we
see that mass is concentrated at much larger energies than phase-space
density. This is a simple reflection of the fact that the amount of
phase-space volume that is associated with a pixel in \figsand{f_df}{f_mass}
increases rapidly with $\avet{E}$. The surprising feature of \fig{f_mass} is
large concentration of mass in the top right corner of the diagram. This
mass is on retrograde orbits around corotation. It lies there because at
these effective energies the isotropic component cannot contribute, and
there is a lack of regular prograde orbits.

We find that 75 per cent of the mass in the box is on orbits that are
confined to be inside corotation. Of these, about a sixth (12 per cent) are
in the isotropic component, half (37 per cent) are prograde regular and a
third (26 per cent of the total in the box) are retrograde regular. Thus
Within corotation, it is prograde orbits that are dominant. We have tried
and failed to build models that place a much larger fractions of mass on the
isotropic component. Inside corotation, chaotic orbits play quite a small
role, although, as \fig{massprof} illustrates, they are indispensable along
the minor axis.  Outside corotation, retrograde orbits dominate the mass
budget. The second column of \tab{mass-tau} gives the mass contained in each
component.

\fig{dmde} gives an overview of the distribution of mass between the various
components by plotting $\d M/\d E$, the mass per unit increment in energy.
Since the DF of the isotropic component is a function $\fe(H)$ of the
Hamiltonian, calculating the corresponding form of $\d M/\d E$ involves
determining the volume in phase space within which both $E(\bw)$ and
$H(\bw)$ lie within specified ranges, and then multiplying this volume by
$\fe$ and integrating over all $H$. Details of this calculation are given in
\App{dmde}. The long-dashed curve in \fig{dmde} shows the resulting curve,
which rises fairly steadily with $E$ until it drops sharply to zero as the
cutoff at $H=\Phi_{\rm eff}^{\rm rim}$ cuts in.

The full curve in \fig{dmde} shows $\d M/\d E$ for the regular component in
the approximation that each orbit contributes mass only to the energy that
is equal to $\avet{E}$, the time-average of $E(\bw)$ along the orbit. Since
$E$ generally does not vary greatly along a regular orbit, this is a good
approximation. Comparing the full and long-dashed curves in \fig{dmde}, we
see that inside corotation the regular and isotropic components make
comparable contributions to the overall mass, with the isotropic component
dominant at the lowest energies, and the regular component mostly dominant
further out. The short-dashed and dotted curves in \fig{dmde} show the
contributions to the regular component from prograde and retrograde orbits,
respectively. Prograde orbits generally contribute more than half the mass.
Around the corotation energy, $E=-0.05$, there is a glaring exception to
this rule, however, as the contributions to $\d M$ from prograde orbits
plunges to zero and retrograde orbits provide all the mass. At slightly
higher energies the relative importance of the two orbit types reverses,
sharply.

\begin{table}
\caption[]{Mass and optical depth towards Baade's Window (BW) and the Galactic 
	Centre (GC). \label{tab:mass-tau} }
\begin{center}
\begin{tabular}{l@{$\;\;$\vline$\;\;$}cc@{$\;\;$}c@{$\quad$}c@{$\;\;$}c}
component &mass&\multicolumn{2}{c}{Sources by comp.}
	       &\multicolumn{2}{c}{Lenses by comp.} \\
	  &    & GC & BW & GC & BW \\ \hline
isotropic &0.12&$1.1\,10^{-5}$&$1.1\,10^{-6}$ & $3.5\,10^{-6}$&$3.6\,10^{-7}$\\
prograde  &0.32&$2.3\,10^{-5}$&$1.3\,10^{-6}$ & $2.9\,10^{-6}$&$5.1\,10^{-7}$\\
retrograde&0.39&$4.6\,10^{-5}$&$2.0\,10^{-6}$ & $3.9\,10^{-6}$&$2.7\,10^{-7}$\\
hot 	  &0.17&$1.3\,10^{-5}$&$1.3\,10^{-6}$ & $9.4\,10^{-6}$&$1.1\,10^{-7}$
\\[1ex]
total 	  &1.00&$-$&$-$ & $2.0\,10^{-5}$&$1.2\,10^{-6}$
\\ \hline
\end{tabular}
\end{center}

The prograde, retrograde, and hot components refer to the regular
orbits with $\avet{L_z}/\avet{(L_z-\avet{L_z})^2}^{1/2}$ smaller than
$-3$, larger than $3$, or in between. The relative masses only refer to the
contributions inside our box of $10\nc\times10\nc\times2.8\kpc$. The left
columns give for the {\em sources\/} drawn from the various
components, while the {\em lenses\/} are taken from the entire model
(inside the box). The rightmost two columns are for sources throughout the
model and lenses in individual components.
\end{table}

\subsection{The microlensing optical depth}
\begin{figure*}
\centerline{
	\epsfxsize=80mm
	\epsfbox[78 370 503 811]{\figdir/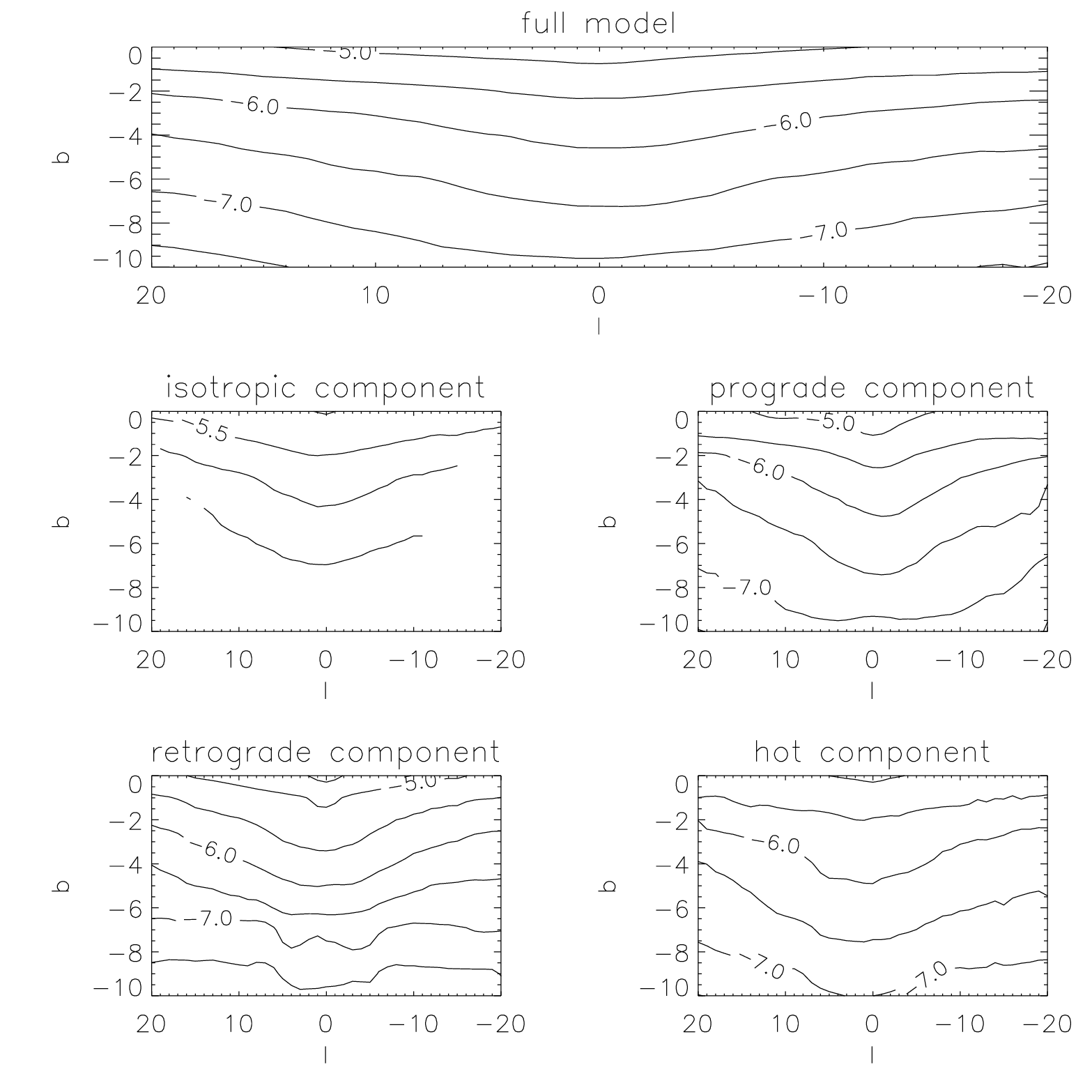} 
	\hspace*{15mm}
	\epsfxsize=80mm
	\epsfbox[78 370 503 811]{\figdir/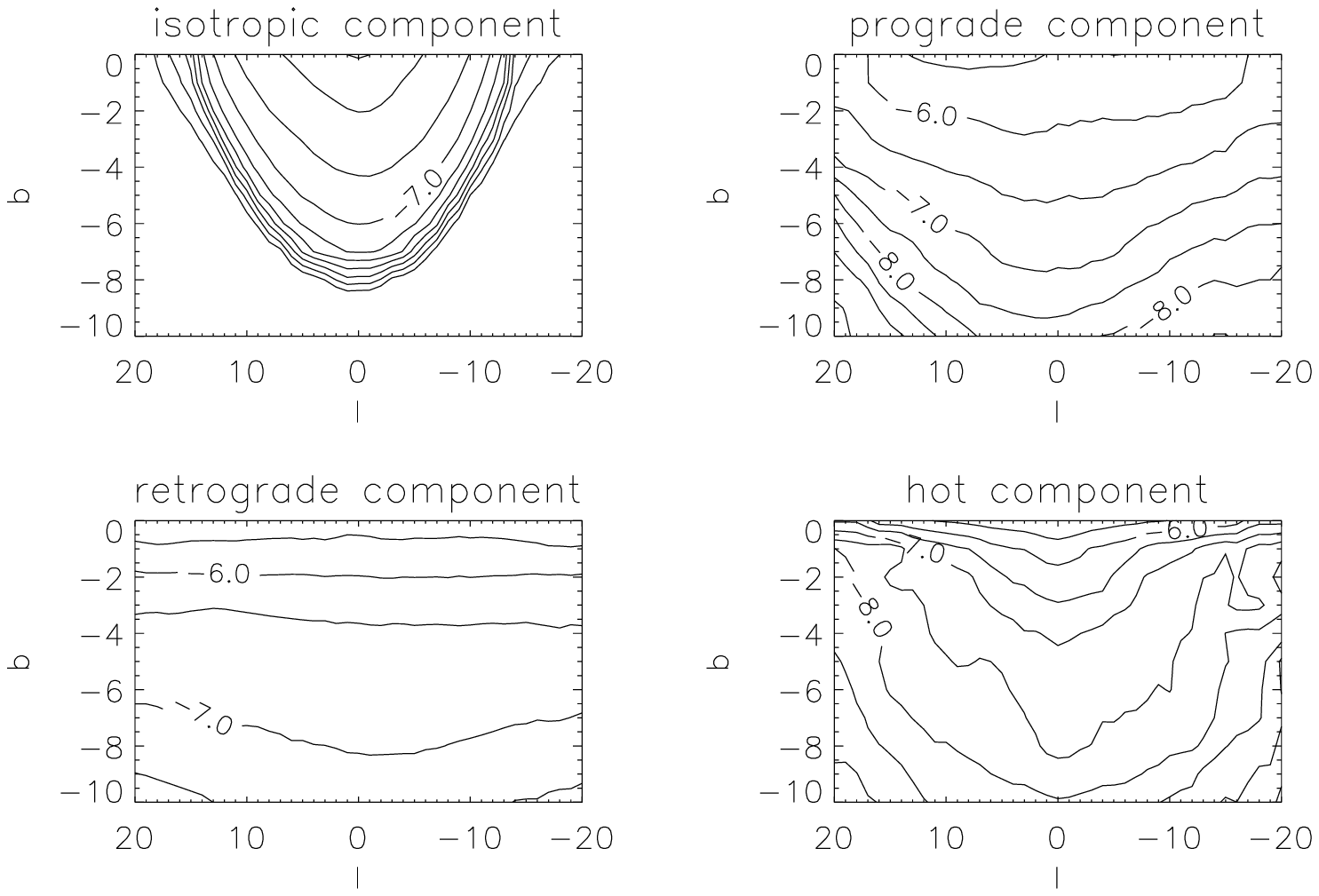}}
\caption[]{Logarithmic (to the base 10) contours of the microlensing optical 
	depth for $\beta\eq0$.
	{\bf Left:} lenses drawn from the full 
	model while the source population differs from panel to panel; this 
	corresponds to the results on the left part of \tab{mass-tau}. 
	{\bf Right:} sources drawn from the full model while the lens population 
	differs from panel to panel.
	Note that the density, and hence the optical depth, of the 
	component with the isotropic DF ends abruptly at corotation.
	\label{fig:tau}}
\end{figure*}

The optical depth for gravitational microlensing, $\tau$, is given by
 \beq
	\tau = {4\upi G\over c^2} { \ds
	 \int\limits_0^\infty\!\d\Ds\,{\Ds^2\,\rhos(\Ds)\over\Ds^{2\beta}}
	 \int\limits_0^{\Ds} \!\d\Dd\,{\Dd(\Ds{-}\Dd)\rhod(\Dd)\over\Ds}
	\!\! \over \ds
	 \int\limits_0^\infty\!\d\Ds\,{\Ds^2\,\rhos(\Ds)\over\Ds^{2\beta}}
	}.
\eeq
Here, the subscripts d and s refer to deflector and source objects,
respectively and $\beta$ parameterizes the efficiency with which stars are
picked up along the line of sight (see e.g., Kiraga \& Paczy\'nski 1994).
The value $\beta\eq0$ is appropriate for sources luminous enough to be
included in a microlensing survey no matter how far away they lie within the
Galaxy, while $\beta\eq1$ is appropriate for less luminous sources that have
a probability of being included in the survey that falls off with
heliocentric distance $s$ as $s^{-1}$. Typically, $\beta\eq0$ is appropriate
for red-clump stars, while $\beta\ap1$ is appropriate for main-sequence
stars.  The top left panel of \fig{tau} shows contours of optical depth
for the case $\beta=0$, while
the bottom entries in the last two columns of \tab{mass-tau} give 
numerical values for the direction  towards the Galactic
centre and  Baade's Window.  These depths are only for microlensing a
source that lies within the box by a lens that also lies within the box.
Hence, before they are compared with observations, these numbers should be
augmented by the contribution from the foreground disk, which for Baade's
Window typically amounts to $0.1$ - $0.3\nc\times 10^{-6}$ (e.g., Evans
1995).  The augmented value for Baade's Window, namely
$1.4\nc\times10^{-6}$, is smaller than the observational value: Udalski et
al.\ (1994) found $\tau\eq(3.3\nc\pm 1.2)\nc\times 10^{-6}$ for clump stars,
while Alcock et al. (1997) found $\tau\eq(3.9^{+1.8}_{-1.2})\nc\times 10^{-6}$
for clump stars near to Baade's Window. This problem was discussed in detail by
Bissantz et al.\ (1997), who examined essentially the input density
distribution.

The four right-hand panels of \fig{tau} and the other entries in the last
two columns of \tab{mass-tau} break the overall optical depth down into
contributions from individual components. Towards the Galactic centre, the
largest contribution comes from lenses in the hot component, while in
Baade's Window this component makes the smallest contribution because of its
high degree of  central concentration. Nearly half the
overall optical depth in Baade's Window comes from lenses in the prograde
component.

The four lower left panels in \fig{tau} and the second
and third columns of \tab{mass-tau} give the optical depth values when the
lenses are drawn from the full model, but the sources are separated by
component. The table shows that these vary by a factor of up to 2 in a given
direction.  Specifically, in Baade's window, sources that move on retrograde
orbits are nearly twice as likely to be lensed as sources that belong to the
isotropic component, and, in fact, have an optical depth that is compatible
with the measurement of Udalski et al.  The origin of this phenomenon is
simple: objects that lie near the outside of the box, but diametrically
opposite to the Sun, inevitably have larger optical depths than objects at
the Galactic centre.  \fig{dmde} shows that the largest radii are dominated
by retrograde orbits, while the isotropic component is concentrated at the
centre. There is a real possibility that the density of clump stars does not
rise towards the Galactic centre as rapidly as the overall stellar mass
density, with the result that their optical depth is anomalously high like
that of the model's retrograde objects.

Barred models can possess asymmetric microlensing maps (Evans 1994, Zhao \&
Mao 1996). Two competing effects can be discerned in \fig{tau}. First, lines
of sight to sources at negative longitudes are longer and pass through more
of the dense central bulge than lines of sight at positive longitudes.  This
is because the sources at negative longitudes are on average further away.
Second, the selection effect, controlled by the parameter $\beta$, can
enhance the optical depth at positive longitudes rather than negative, as
here the sources appear to be brighter. These two effects tend to work
against each other, so the final asymmetry of the map is a subtle matter.
For example, in the  panel on the extreme right of the middle row of
\fig{tau}, which is for lenses that lie in the prograde component and 
sources that are drawn from the full dynamical model, we see enhanced
optical depths at negative longitudes, but only outside the plane. Within
the plane, any interpretation is difficult because this is where our
dynamical model is least reliable. By contrast, when the sources lie in the
retrograde component near the Galactic plane (extreme left panel of
bottom row of \fig{tau}), the asymmetry goes the other way.  Here, the main
contribution is from sources at the far end of the Galaxy outside corotation,
and now the optical depth is enhanced if the lenses lie roughly halfway
between observer and source. This favours positive longitudes, as there is a
ready supply of such lenses in the foreground bar. 

From the top-left panel of \fig{tau} we see that the optical depth is always
larger at negative longitudes when $|b|\nc>1\degr$. The coming decade is
likely to see the gradient of microlensing optical depth with Galactic
longitude and latitude measured, at least at certain spots in the Bulge. For
example, the third panel from the left in the bottom row of \fig{tau} shows
that when the lenses are drawn from the retrograde component, the gradient
with respect to Galactic longitude near the plane is negligible, although
the latitudinal gradient remains steep. At on-axis ($b\nc\sim0\degr$) fields
with $\ell\nc>10\degr$, the optical depth is mainly provided by the
retrograde component. Hence, gradient information, by giving an indication
of the shape of the contours in the microlensing maps, may provide clues as
to which components are providing the largest numbers of sources and lenses.

\section{CONCLUSIONS}

We have shown how Schwarzschild's method can be used to produce models with
known DFs. Previously, models constructed by Schwarzschild's (1979) method
have had known DFs only when the DF depended only on classical integrals
(e.g., Cretton et al.\ 1999).  The great merit of Schwarzschild's method is
precisely its ability to handle the general case in which non-classical
integrals are important, so our algorithm for determining the DF
of a general model must be counted a significant advance.

One advantage of being able to determine the DF of a Schwarzschild model is
that one can then combine a crude approximation to the galaxy with a DF
that depends only on classical integrals, with a more general DF obtained by
Schwarzschild's method. In this approach, only the difference between the
true DF and the classical DF need be reproduced with orbits. The resulting
model will have higher resolution in real- and velocity-space, and be easier
to interpret physically than one constructed by the classical Schwarzschild
technique. 

A key point is that when a classical DF is used in the construction of a
Schwarzschild model, the weights of orbits can be negative because the
orbits may be subtracted from the underlying classical
component, subject only to the constraint that the total phase-space density
is non-negative. When regular orbits are assigned negative weights, these
orbits are important because they are excluding stars on chaotic orbits from
certain regular parts of phase space.

We have illustrated these new techniques by using them to construct a model
of the central kiloparsecs of the Milky Way. This problem is less well
suited to the new techniques than the classical problems of modelling
axisymmetric systems and triaxial systems with negligible figure rotation
because there is only one classical integral, the Jacobi energy, and this
integral is not confining for the more energetic stars.  Moreover, the phase
space of the inner Galaxy is more chaotic than regular, and Schwarzschild's
technique works best in a highly regular phase space.  From the fact that we
have succeeded in constructing a reasonable model of the inner Galaxy
notwithstanding these difficulties, we infer that the new techniques would
solve easier classical problems with some facility.

Our new Galaxy
model reproduces essentially all of the available density and kinematic
data within corotation (${\sim}\,3.6\kpc$). In particular, the
three-dimensional bar density of Binney et al.\ (1997) is faithfully
reproduced in the inner parts. At corotation, our dynamical model does not
reproduce the over-densities on the minor axis found by Binney et al.\ --
but these are probably artifacts of the deprojection algorithm, which
assumes that the underlying model is eightfold symmetric and must
misrepresent spiral features. The kinematic data in Baade's Window \wBaa\
and in the field studied by Minniti et al.\ (1992) at \wMia\ are all fitted
to within the observational uncertainties. The model does not fit data for  the outer
window of Minniti et al.\ (1992) at \wMib\ accurately in that it
has a higher line-of-sight dispersion and a lower streaming velocity than
the data. This mismatch may be caused by disk contamination in the Minniti
et al.\ sample or by too low an assumed value for the mean extinction in our
model.

In addition to fitting the initially prescribed data, the model furnishes
many predictions. We have tested a few of these predictions against data in
the literature and find that they are consistent with measurements by Blum
et al.\ (1994) at \wBla\ and \wBlb, and by Tiede \& Terndrup (1997) at \wTT.
For several well studied fields we predict values for quantities, such as
proper-motion dispersions, that have not yet been measured.  We find that in
central fields, such as Baade's Window, the principal axes of the
velocity-dispersion tensor tend to be approximately aligned with the line of
sight. Hence, these are not good locations to probe for kinematic signatures
of triaxiality.

The most puzzling aspect of our model is its reliance on retrograde orbits
in a narrow band of energies, that corresponds to radii around $3\kpc$. This
fact has an observationally testable consequence: at \wMia\ the dispersion
tensor should be extremely anisotropic, with its longest axis aligned nearly
with the $\bel$ direction, because samples should contain roughly equal
numbers of prograde and retrograde stars, on roughly circular orbits. Is
this counter-intuitive prediction an artifact of our model or a robust
prediction? It is a consequence of two inputs: (i) the density profile of
Binney et al., and (ii) the decision to exclude from the orbit library
orbits with Liapunov times shorter than five bar rotation periods. 

The density distribution of Binney et al.\ is very uncertain at the relevant
radii by virtue of a combination of the effects of spiral structure and
obscuration. Photometry of external barred galaxies, such as NGC 1300
(Elmegreen et al.\ 1996) suggests that the Binney et al.\ profile may be
significantly in error, and that the density should drop steeply inside
corotation to a lower, approximately level value that extends throughout the
highly chaotic region around corotation. An approximately constant density
around corotation could be provided by prograde orbits that have apocentres
well outside corotation. By specifying a density gradient around corotation,
we have obliged the fitting algorithm to employ orbits with apocentres at
smaller radii, and the only available regular orbits are retrograde.
Alternatively, we may have been too conservative in our criterion for
excluding orbits as chaotic: an orbit that has a short Liapunov time may
never the less remain trapped for a substantial fraction of a bar rotation
period. In a subsequent study it would be interesting to use the spectral
approach to determining orbital regularity (Binney \& Spergel 1984;
Carpintero \& Aguilar, 1998) instead of Liapunov exponents, which are
expensive to compute and may be more misleading in cases of mild
stochasticity.  Moreover, even though half of all disk galaxies have bars,
it is not clear that any individual bar lives for a Hubble time; bars may
dissolve and reform on a shorter timescale.

We have calculated microlensing maps for our model. These show that the
overall microlensing optical depth of luminous sources in Baade's window is
at least a factor 2 smaller than the observations require. They show also
that sources in different components have optical depths that differ by up
to a factor 2. Hence, it is important to characterize the populations to
which stars that are known to have been lensed belong, and to understand the
mix of populations that characterizes the stars that are regularly monitored
for microlensing events. In principle, our model allows one to determine the
distribution of durations of the lensing events in any field, and we plan to
present such distributions shortly.

It would be useful to extend the work in this paper by building dynamical models
for different bar morphologies, viewing angles and pattern speeds. However, our
experiments have convinced us that there is very considerable freedom to
reproduce the existing data by superposition of orbits. This means that
restricting the viewing angle from the stellar kinematics alone will be
challenging.

To meet this challenge, more kinematic data in the inner Galaxy are urgently
needed. It may well be possible to extract more information by studying the
variation of kinematic quantities with distance along the line of sight
using well-defined selection functions. For external galaxies, where all
stars are at roughly the same distance from the observer, it makes sense to
record a single number for a kinematic quantity like the velocity dispersion
along the line of sight. In studies of the Milky Way, this approach fails to
exploit the full richness of information that is available to us. It would
be more fruitful to calculate the kinematic quantities for stars in the
sample with, for example, magnitudes between $m_{\rm max}$ and $m_{\rm
min}\eq m_{\rm max}\mi1$.  As we have shown for the dynamical model, there
are interesting and useful variations of kinematic quantities with $m_{\rm
max}$.  These will prove invaluable in elucidating the structure of the
bar, since the photometry together with the available line-of-sight
dispersions and streaming velocities are not enough to prescribe the bar
uniquely.

\section*{Acknowledgments}
RMH and WD acknowledge financial support from PPARC, while NWE is supported by
the Royal Society. We thank Ortwin Gerhard and Hongsheng Zhao for helpful
discussions.

\begin{appendix}	
\section{NUMERICAL ALGORITHM FOR THE POTENTIAL}
This appendix sketches the calculation of the potential for the Binney et al.\
(1997) bar density.  A straightforward expansion in spherical harmonics
converges rather slowly, especially in the flat disk component.  However, this
problem can be overcome by a technique due to Kuijken \& Dubinski (1994).  
Given a disk with density $\rho\eq f(R)h(z)$, the potential is written as
$\Phi\eq\tilde{\Phi} + f(r)H(z)$, where $r$ denotes spherical radius and $H(z)$
is uniquely determined by $H^{\prime\prime}(z)\equiv h(z)$ and $H(0)\eq
H^\prime(0) \eq0$ (primes denoting derivatives). Inserting this into Poisson's
equation gives
 \beq \bea{rcl}
	\ds        {\nabla^2\tilde{\Phi}\over 4\upi G} & = 
	\ds \big[f(R)-f(r)\big]h(z)&\ds - f^{\prime\prime}\!(r)H(z)	\\
	& & \ds - {2f^\prime\!(r)\over r} \big[H(z)+zH^\prime\!(z)\big].
\eea \eeq
 The expression on the right hand side is zero at $z\eq0$. The mass density
generating $\tilde{\Phi}$ is not strongly flattened, so that it can be
economically evaluated by, e.g., multipole expansion. Here, this technique is
used for the two sub-disks. The subsequently evaluated potential $\tilde{\Phi}$
and the modified density is then expanded in spherical harmonics up to order
$l_{\rm max}\eq64$. The sharp truncation of this expansion at $l_{\rm max}$
causes unphysical ringing of the resulting density and in order suppress this we
have tapered the density expansion by multiplying it with $\exp(-[l/32]^2)$.
Finally, the potential and resulting forces have been computed on a
pseudo-Cartesian $101\times101\times161$ grid of size
$20\kpc\times20\kpc\times18\kpc$. The points are linear in $2\ln(1+x/2)$,
$2\ln(1+y/2)$, and $\ln(1+2z)/2$. Potential and forces are then evaluated via a
fifth order three dimensional spline. At each grid point, this spline gives the
stored values of potential and its derivatives. The forces have everywhere
continuous first and second derivatives. In particular, the interpolated forces
agree identically with derivatives of the interpolated potential, as is
necessary if the Jacobi energy is to be accurately conserved along numerically
integrated orbits.  The evaluation of potential and forces is quick once the
spline coefficients have been pre-computed.

\section{CALCULATION OF THE SECOND VELOCITY MOMENTS}
This appendix briefly sketches how to calculate the second moments for the
regular and isotropic parts of the DF in turn.

\subsection{Regular part of DF}
To compute a velocity moment projected along the line of sight towards
$(\ell_0,\,b_0)$ with the formalism of Section~\ref{sec:formal}, we define
\beq
	\widehat{\Pi}(\bw) \equiv \delta(\ell-\ell_0)\,\delta(b-b_0)
		\,\eps(s)\,g(\bw).
\eeq
 where $g(\bw)$ is determined by the velocity moment in question. For example,
to obtain $\vlos$, one takes $g\eq\bp\cdot\hat{\bs}$, where $\hat{\bs}$ is the
unit vector in direction of the line of sight. With this choice, one obtains for
the projected moment using \eqnsto{Pioff}{Pisum}
 \beq \bea{rcl}
	\Pi[\fr] &= \ds {1\over\Ns} \sum_i^{\Nr}\wr_i\,
		\lim_{T\to\infty}&\ds{1\over T}
		\int_0^T\!\d t\,\eps\big(s_i(t)\big)
		\,g\big(\bw_i(t)\big)	\\ & & \ds \times 
		\,\tilde{\delta}\big(\ell_i(t){-}\ell_0\big)
		\,\tilde{\delta}\big(b_i(t){-}b_0\big)
\eea \eeq
 Here, $\tilde\delta(x)$ denotes a ``$\delta$-function'' of some finite width,
which in Baade's Window, for example, is 30\arcmin. The reason that we cannot
use strict $\delta$-functions is that the probability for any single orbit to
hit the line of sight in our finite integration interval of 200 dynamical times
is zero. That is, the need for the finite-width functions $\tilde\delta$ arises
because we replaced (i) the integral over action-space by a finite sum (by
virtue of the Monte-Carlo method), and (ii) the integral over angles by a
finite, rather than infinite, time integral.

\subsection{Isotropic part of DF}
 Computing moments of the isotropic part of the DF $\fe$ can be done directly
since $\fe$ is explicitly given as a function of Jacobi energy $\EJ$ and
therefore phase-space coordinates $\bw$. The projected luminosity density is
 \beq\label{nu}
	\nu(\ell,b) = 4\upi\int\d s\,s^2\,\eps(s)\int
	\d v\,v^2\,f\big(\br(s,\ell,b),\bv\big).
\eeq
 The second integral can be performed analytically for the second order splines
$B_i(\EJ)$ that constitute the building blocks of $\fe$ -- see \eqn{equ_rhoc}.
The integration over $s$ is done numerically.

Since $\EJ$ depends only on the modulus of the velocity (and not $\bv$ itself),
the first moments $\int\d\bv\,\bv \fe$ vanish and one obtains $\ave{v_i}\eq-
v_{\odot i}$, i.e.\ the solar motions, while
 \beq
	\ave{\mu_i} = - {v_{\odot i}\over\nu(\ell,b)}
		\int \d s\, s^2\, \eps(s) {\rhoi(\br)\over s}
\eeq
 which is just the solar reflex motion $v_{\odot i}/s$ weighted by the density
along the line of sight. For the second velocity moments one finds ($p\id|\bp|$)
\beq
	\nu\sigma_{ij}^2 = \delta_{ij}\,{4\upi\over3}
		\int\d s\,s^2\,\eps(s) \int\d p\; p^4 
		\fe\big(\br(s,\ell,b),p\big),
\eeq
where the inner integral can be performed analytically by the aid of
\beq
	\int\d p\, p^4\, B_i(\EJ) =
		\frac{8\sqrt{2}\,k_i}{35\,\Delta\EJ} 
		\left[D_{i-1}^7-2 D_i^7+D_{i+1}^7\right]
\eeq
 with $D_i^2\id\EJ_i-\Peff(\br)$. The outer integral is performed numerically.
For dispersions involving proper motions rather than velocities, one has to
change the integrand by factors of $s^{-1}$.

\section{DIFFERENTIAL MASS IN THE ISOTROPIC COMPONENT} \label{app:dmde}

We want to calculate the differential mass $\d M/\d E$, where $E$ denotes the
energy, for the isotropic part $\fe(\EJ)$ of the distribution function. Let
us consider a $\delta$-function at $\EJ\eq\EJ_0$, then its contribution
to $\d M/\d E$ at $E\eq E_0$ is simply the volume of the four-dimensional
cross-section of the phase-space hyper-planes at $\EJ\eq\EJ_0$ and $E\eq E_0$
\beq \label{gee}
	g(E_0|\EJ_0) \equiv \int\d^6\bw\;
		\delta(\EJ-\EJ_0)\;\delta(E-E_0).
\eeq
Then
\beq
	{\d M\over\d E} = \Mb \int\d\EJ\, g(E|\EJ)\, \fe(\EJ),
\eeq
i.e.\ $g$ is the density of states at given $\EJ$. With $\d^3\bp\eq 2\upi R^{-1}
\d E\,\d L_z$ (after integrating out one angle in $\bp$-space) and $\EJ\eq H-
\Omega L_z$ \beqn{defham} (\ref{gee}) becomes
\ben
	g(E_0|\EJ) &=& {2\upi\over \Omega} \int {\d^3\!\br\over R} 
		\int_{\Phi(\br)}^{\Phi(\infty)}\d E\;\delta(E-E_0)
		\nonumber \\ & & \times
		\int_{-R\sqrt{2(E-\Phi)}}^{+R\sqrt{2(E-\Phi)}}
		\d L_z\;\delta\left(L_z-{E{-}\EJ\over\Omega}\right)
		\nonumber \\
	&=& {2\upi\over\Omega}\int\d\phi\int\d z\;(R_2-R_1).
\een
Here, $R_{1/2}(E_0,\EJ,z,\phi)$ are the roots of 
\beq	
	(E_0 - \EJ)^2= 2R^2\Omega^2 \big(E_0 - \Phi(R,z,\phi)\big),
\eeq	
more precisely, the minimum of these roots and the corotation radius -- recall 
that the isotropic component lives only inside corotation.

\end{appendix}
\end{document}